\documentclass[preprint,10pt,authoryear,a4paper,twocolumn]{elsarticle}
\usepackage{setspace,graphicx,epstopdf,amsmath,amsfonts,amssymb,amsthm}
\usepackage{marginnote,datetime,enumitem,subfigure,rotating,fancyvrb}
\usepackage{hyperref,float}
\usepackage{rotating}
\usepackage{natbib}
\usepackage{bm}
\usepackage{graphicx}
\usepackage{pdflscape}
\usepackage{xcolor}
\usepackage{amsmath}
\usepackage{mathtools}
\usepackage{cuted}
\usepackage{lipsum, color}
\usepackage{graphics}
\hypersetup{
    colorlinks,
    linkcolor={red!50!green},
    citecolor={red!50!red},
    urlcolor={red!80!blue}
}

\usdate 

\usepackage{listings}
\usepackage{color}

\definecolor{dkgreen}{rgb}{0,0.6,0}
\definecolor{gray}{rgb}{0.5,0.5,0.5}
\definecolor{mauve}{rgb}{0.58,0,0.82}

\usepackage{graphicx}
\usepackage{amssymb,amsfonts,amsmath}
\usepackage[margin=2.5cm]{geometry}
\usepackage{hyperref}
\graphicspath{{./}}
\usepackage{color}

\journal{}

\begin{document}

\setlist{noitemsep}  

\title{\bf Lagrange regularisation approach to compare nested data sets and determine objectively financial bubbles' inceptions}
\author{G. Demos ${\dag}$\footnote{$^\ast$Corresponding author: \url{gdemos@ethz.ch}}, D. Sornette$^{\dag \natural}$\\
$\dag$ {\footnotesize {\it ETH Z\"{u}rich, Dept. of Management, Technology and Economics, Z\"{u}rich, Switzerland.}}\\
$\natural$ {\footnotesize {\it Swiss Finance Institute, c/o University of Geneva, Geneva, Switzerland.}}
}

\date{}              


\maketitle

\centerline{\bf ABSTRACT}
{\scriptsize
Inspired by the question of identifying the start time $\tau$ of financial bubbles, we
address the calibration of time series in which the inception of the latest regime of interest is unknown.
By taking into account the tendency of a given model to overfit data, we introduce
the Lagrange regularisation of the normalised sum of the squared residuals, $\chi^{2}_{np}(\bm{\Phi})$, 
to endogenously detect the optimal fitting window size := $w* \in [\tau:\bar{t}_2]$ 
that should be used for calibration purposes for a fixed pseudo present time $\bar{t}_2$. 
The performance of the Lagrange regularisation of  $\chi^{2}_{np}(\bm{\Phi})$ defined as $\chi^{2}_{\lambda}(\bm{\Phi})$ is
exemplified on a simple Linear Regression problem with a change point and compared 
against the Residual Sum of Squares (RSS) := $\chi^{2}(\bm{\Phi})$ and RSS/(N-p):= $\chi^{2}_{np}(\bm{\Phi})$, 
where $N$ is the sample size and p is the number of degrees of freedom.  Applied to 
synthetic models of financial bubbles with a well-defined transition regime and to a number
of financial time series (US S\&P500, Brazil IBovespa and China SSEC Indices), 
the Lagrange regularisation of $\chi^{2}_{\lambda}(\bm{\Phi})$ is found
to provide well-defined reasonable determinations of the starting times 
for major bubbles such as the bubbles ending with the 1987 Black-Monday, 
the 2008 Sub-prime crisis and minor speculative bubbles on other Indexes, without any further exogenous information. 
It thus allows one to endogenise the determination of the beginning time of bubbles,
a problem that had not received previously a systematic objective solution.}

  \noindent    


\onehalfspacing
\section{Introduction}
There is an inverse relationship between the tendency of a model to overfit data and the sample size used. 
In other words, the smaller the data sample size, the larger the number of degrees of freedom,
the larger is the possibility of overfit \citep{overfittt}. Due this characteristic feature, one cannot compare directly
goodness-of-fit metrics, such as the Residual Sum of Squares (RSS) := $\chi^{2}(\bm{\Phi})$ or its normalized version RSS/(N-p) := $\chi^{2}_{np}(\bm{\Phi})$, of 
statistical models over unequal sized samples for a given parametrisation $\Phi$. Here, $N$ denotes the sample size while $p$ is the number of degrees of freedom of a model.
This is particularly problematic when one is specifically interested in selecting 
the optimal sub-sample of a dataset to calibrate a model.
This is a common problem when calibrating time series,  when the model is only valid
in a specific time window, which is unknown a priori. Our motivation stems from the
question of determining the beginning of a financial bubble, but this question is more generally
applicable to time series exhibiting regime shifts that one is interested in localising precisely.

In the literature, there are solutions for proper model selection such as the Lasso \citep{lasso} and Ridge regressions \citep{ridgeReg}, where the cost function contains an additional penalisation for large values of the estimated parameters. Well-known metrics such as the AIC and BIC are also standard tools for quantifying goodness-of-fit of different models \citep{akaike} and 
for selecting the one with the best compromise between goodness-of-fit and complexity.
However, results stemming from these methodologies are only comparable within the same data set.

There seems to be a gap in the literature about the proper procedure one should follow when 
comparing goodness-of-fit metrics of a model calibrated to different batches of a given data set. In order to fill this gap, we propose a novel metric for calibrating endogenised end points and compare nested data sets.  The method empirically computes the tendency of a model to overfit a data set via what we term the ``{\it Lagrange regulariser term}'' $\lambda$. Once $\lambda$ has been estimated empirically, the cost function can be corrected accordingly as a function of sample size, giving the
Lagrange regularisation of $\chi^{2}_{np}(\bm{\Phi})$. As the number of data points
or the window beginning- or end-point is now endogeneised, 
the optimal sample length can then be determined.
We empirically test the performance of the Lagrange regularisation of $\chi^{2}_{np}(\bm{\Phi})$, which defined
$\chi^{2}_{\lambda}(\bm{\Phi})$ as the regularised Residual Sum of Squares, in comparison with
the naive $\chi^{2}(\bm{\Phi})$ and $\chi^{2}_{np}(\bm{\Phi})$ itself using both linear and non-linear models as well as synthetic and real-world time-series.

This  paper is structured as follows. Section~(\ref{method}) explains the motivation behind the proposed Lagrange regularising term. Moreover, we provide details of the derivation of $\lambda$ as well as the analytical expression for computing the tendency of a model to overfit data. In Section~(\ref{sec3}), we make use of a simple OLS regression to test the empirical performance of the Lagrange regularisation of  $\chi^{2}_{np}(\bm{\Phi})$ on the problem of optimal sub-sample selection. Section~(\ref{sec4}) shows how the regulariser can be used alongside with the LPPLS model of financial bubbles in order to diagnose the beginning of financial bubbles. Empirical findings are given in Sec.~(\ref{sec5}) and Section~(\ref{last})  concludes.

\section{Formulation of calibration with varying window sizes: How to endogenize $t_1$ and make different window sizes comparable}\label{method}

Let us consider the normalised mean-squared residuals, defined as the sum of squares
of the residuals divided by the number $t_2-t_1$ of points in the sum corrected by the number
of degrees of freedom $p$ of the model, 
\begin{align}
\chi^{2}_{np}(\bm{\Phi}) := \frac{1}{(t_{2}-t_{1})-p} \sum_{i=t_{1}}^{t_{2}} r_{i}(\bm{\Phi})^{2}~,
\label{1}
\end{align}
~~~{\rm with}~
\begin{align}
 r_{i}(\bm{\Phi}) = y_i^{data} - y_i^{model}(\bm{\Phi})~,
\end{align}
where $\bm{\Phi}$ denotes the set of model parameters to fit including a priori the
left end point $t_1$ of the calibration window. The term $y_{i}^{model}(\bm{\Phi})$ corresponds to the theoretical 
model and $y_{i}^{data}$ is the empirical value of the time-series at time $i$.
 
For a fixed right end point $t_2$ of the calibration window, we are interested in comparing
the results of the fit of the model to the empirical data for various left end points $t_1$
of the calibration window. The standard approach assumes a fixed  calibration window $[t_1, t_2]$
with $N = t_2-t_1 +1$ data points. In order to relate the two problems, we consider 
the minimisation of $\chi^{2}_{np}(\bm{\Phi})$ at fixed $t_2-t_1$ (for a fixed $t_2$) as
minimising a general problem involving $t_1$ as a fit parameter augmented by the condition 
that $t_2-t_1 +1=N$ is fixed.
This reads
\begin{equation}
{\text{\rm Min}}~~ \chi^{2}_\lambda(\bm{\Phi}) ~,
\end{equation}
~{\rm with} ~
\begin{equation}
 \chi^{2}_\lambda(\bm{\Phi}) := \frac{1}{(t_{2}-t_{1}) - p} \sum_{i=t_{1}}^{t_{2}} r_{i}(\bm{\Phi})^{2}  + \lambda (t_2-t_1)~,
\label{tg2efqtrbg}
\end{equation}
where we have introduced the Lagrange parameter $\lambda$, which is conjugate to the constraint $t_2-t_1 +1=N$.
Once the parameters $\bm{\Phi}$ are determined, $\lambda$ is obtained 
by the condition that the constraint $t_2-t_1 +1=N$ is verified.

Since data points are discrete, the minimisation of (\ref{tg2efqtrbg}) with respect to $t_1$ reads

\begin{strip}
\begin{align}
0 &=& \chi^{2}_\lambda(\bm{\Phi})(t_1 +1) - \chi^{2}_\lambda(\bm{\Phi})(t_1)  = 
\frac{1}{(t_{2}-t_{1}-p-1)} \sum_{i=t_{1}+1}^{t_{2}} r_{i}(\bm{\Phi})^{2} 
- \frac{1}{t_{2}-t_{1}-p} \sum_{i=t_{1}}^{t_{2}} r_{i}(\bm{\Phi})^{2}  - \lambda ~  \nonumber \\
& = & \frac{1}{t_{2}-t_1-p} \left(1 + {1 \over t_2 - t_1-p} + {\cal O}
 \left({1 \over (t_2 - t_1-p)^2}\right) \right) \sum_{i=t_{1}+1}^{t_{2}} r_{i}(\bm{\Phi}) ^{2} 
- \frac{1}{t_{2}-t_{1}-p} \sum_{i=t_{1}}^{t_{2}} r_{i}(\bm{\Phi})^{2}  - \lambda ~, \nonumber \\
&=& - {1 \over t_2 - t_1-p} r_{t_1}(\bm{\Phi})^{2} \left(1 +  {\cal O} \left({1 \over t_2 - t_1-p}\right)\right) + \frac{1}{t_{2}-t_{1}-p} \chi^{2}(\bm{\Phi})
 \left(1 +  {\cal O} \left({1 \over t_2 - t_1-p}\right)\right)    - \lambda ~.
\end{align}
\end{strip}
Neglecting the small terms ${\cal O} \left({1 \over t_2 - t_1-p}\right)$ leads to 
\begin{equation}
 \chi^{2}_{\lambda}(\bm{\Phi}) =  r_{t_1}(\bm{\Phi})^{2} + \lambda (t_2-t_1-p)~.
 \label{wtrhrygq}
\end{equation}

Expression (\ref{wtrhrygq}) has the following implications.
Consider the case where all squared terms $r_{i}(\bm{\Phi})^{2}$ in the sum (\ref{1}) defining $\chi^{2}_{\lambda}(\bm{\Phi})$
are approximately the same and independent of $t_1$, which occurs when the residuals 
are thin-tailed distributed and the model is well specified. Then, we have
\begin{equation}
r_{i}(\bm{\Phi})^{2} \approx r^2~,   ~~\forall i~, ~ {\rm including}~  r_{t_1}(\bm{\Phi})^{2}=r^2~,
\label{thythwgq}
\end{equation}
and thus 
\begin{equation}
\chi^{2}_{np}(\bm{\Phi}) \approx r^2~.
 \label{wtrgt1hrygq}
\end{equation}
Expressing  (\ref{wtrhrygq})  with the estimation (\ref{thythwgq}) yields
\begin{equation}
\chi^{2}_{\lambda}(\bm{\Phi}) \approx r^2 + \lambda (t_2-t_1-p)~.
\end{equation}
Comparing with (\ref{wtrgt1hrygq}), this suggests that varying $t_1$
is expected in general to introduce a linear bias of the normalised
sum $\chi^{2}_{np}(\bm{\Phi})$ of squares of the residuals, which is proportional
to the size of the calibration window (up to the small correction by
the number $p$ of degrees of freedom of the model). If we want to compare the calibrations over different
window sizes, we need to correct for this bias. 

More specifically, rather than fixing the window size $t_2-t_1 +1=N$, we want to determine the `best' $t_1$,
thus comparing calibrations for varying window sizes, for a fixed right end point $t_2$.
As a consequence, the Lagrange multiplier $\lambda$ is no more fixed to ensure that the constraint 
$t_2-t_1 +1=N$ holds, but now quantifies the average bias or ``cost'' associated with changing the 
window sizes. This bias is appreciable for small data sample sizes. It vanishes asymptotically
as $N\rightarrow \infty$, i.e. ${\rm lim}_{N\rightarrow \infty}\lambda=0$.

In statistical physics, this is analogous to the change from the canonical to the
grand canonical ensemble, where the condition of a fixed number of particles (fixed number
of points in a fixed window size) is relaxed to a varying number of particles with an energy
cost per particle determined by the chemical potential (the Lagrange parameter $\lambda$) \citep{ensemble}. 
It is well-known that the canonical ensemble is recovered from the grand canonical ensemble
by fixing the chemical potential (Lagrange multiplier) so that the number of particles is equal
to the imposed constraints. Idem here. 

How to determine the crucial Lagrange parameter $\lambda$?
We propose an empirical approach. When plotting  $\chi^{2}_{np}(\bm{\Phi})$ 
as a function of $t_1$ for various instances, we observe that 
a linearly decreasing function of $t_1$ provides a good approximation of it,
as predicted by (\ref{wtrhrygq}) (for $\lambda>0$). The slope 
can then be interpreted as quantifying the average bias of the 
scaled goodness-of-fit $\chi^{2}_{np}(\bm{\Phi})$ due to the reduced number
of data points as $t_1$ is increased. This average bias 
is clearly dependent on the data and of the model used to calibrate it. 
We can thus interpret the average linear trend observed empirically as determining
the effective Lagrange regulariser term $\lambda$ that quantifies the impact on 
the goodness-of-fit resulting from the addition of data points in the calibration, given the specific realisation of the data and
the model to calibrate.
Thus, to make all the calibrations performed for different $t_1$ comparable for the determination of the
optimal window size, we propose to correct expression (\ref{1}) by subtracting 
the term $ \lambda (t_2-t_1)$
from the normalised sum of squared residuals  $\chi^{2}_{np}(\bm{\Phi})$ given by Eq. (\ref{1}), where $\lambda$
is estimated empirically as the large scale linear trend. Here, we omit the $p$ correction since it leads to a constant
translation for a given model with given number of degrees of freedom.
Such a large scale linear trend
of $\chi^{2}_{np}(\bm{\Phi})$ as a function of $t_1$ has been reported for a number of 
financial bubble calibrations in  \citep{DemosSornette2015}.
Our proposed procedure thus amounts simply to detrend $\chi^{2}_{np}(\bm{\Phi})$,
which has the effect of making more pronounced the minima of $\chi^{2}_{np}(\bm{\Phi})$, 
as we shall see below for different models. 

To summarise, endogenising $t_1$ in the set of parameters to calibrate requires to minimize
\begin{align}
\chi^{2}_{\lambda}(\bm{\Phi}) &= \chi^{2}_{np}(\bm{\Phi}) - \lambda (t_2 - t_1) \\
& = \frac{1}{(t_{2}-t_{1}) - p} \sum_{i=t_{1}}^{t_{2}} r_{i}(\bm{\Phi})^{2}  - \lambda (t_2 - t_1)~,
\end{align}
~{\rm with,}~
\begin{equation}
 r_{i}(\bm{\Phi}) = y_i^{data} - y_i^{model}(\bm{\Phi})~,
\label{2yh2y2}
\end{equation}
where $\lambda$ is determined empirically so that $\chi^{2}_{np}(\bm{\Phi}) - \lambda (t_2 - t_1)$
has zero drift as a function of $t_1$ over the set of scanned values. The obtained empirical value of $\lambda$ can be used as a diagnostic parameter quantifying the tendency of the model to over-fit the data. We can 
thus also refer to $\lambda$ as the ``overfit measure''. When it is large, the goodness-of-fit $\chi^{2}(\bm{\Phi})$
changes a lot with the number of data points, indicating a poor overall ability of the model to account for the data.
\cite{DemosSornette2015} observed other cases where $\chi^{2}(\bm{\Phi})$ is constant as a function of $t_1$
(corresponding to a vanishing $\lambda$),
which can be interpreted in a regime where the model fits robustly the data, ``synchronizing'' on
its characteristic features in a way mostly independent of the number of data points.

\section{Application of the Lagrange regularisation method to a simple linear-regression problem}\label{sec3}

Consider the following linear model:
\begin{align}
\bm{Y} = \beta \bm{X} + \bm{\varepsilon},\label{ols}
\end{align} 
with explanatory variable of length $(N \times 1)$ denoted by $X = \{x_1,  x_2, \ldots, x_N\}$, regressand $Y = \{y_1, y_2, \ldots, y_N \}$ and error vector $\bm{\varepsilon} \sim \mathcal{N}(0, \sigma^2)$. Bold variables denote either matrices or vectors. Fitting Eq.~(\ref{ols}) to a given data set $Y^{data}$ consists on solving the quadratic minimisation problem
\begin{align}
\hat{\beta} = \underset{\beta}{arg\, min}~ \chi^{2}(\bm{\Phi}),
\label{nbwgqw}
\end{align}
where $\bm{\Phi}$ are parameters to be estimated and the objective function $\chi^2(\bm{\Phi})$ is given by
\begin{align}
\chi^2(\bm{\Phi}) &= \sum_{i=1}^N|Y_{i}^{data} - (Y_i^{model} - \beta X_i)|^{2}\\
 &= ||\bm{Y}^{data} - (\bm{Y}^{model} - \bm{\beta X})||.\label{costL}
\end{align}

The solution of Eq.~(\ref{nbwgqw}) with (\ref{costL}) for a given data set of length $N$ reads
\begin{align}
\hat{\beta} &= (\bm{X}^{\prime}\bm{X})^{-1} \bm{X}^{\prime}\bm{Y}.
\end{align}

Let $w^{*} \subseteq Y^{data}$ and have length $\leq$ N. $w^{*} \in [\tau:\bar{t}_2]$ thus denotes the optimal window size one should use for fitting a model into a data set of length N for a fixed end point := $t_2$ and an optimal starting point := $\tau$.  

In order to show how the goodness-of-fit metric $\chi^2(\bm{\Phi})$ fails to flag the optimal $\tau$-portion of the data set where 
the regime of interest exists and how delicate is $\chi^2_{np}(\bm{\Phi})$ for diagnosing the true value of 
the transition time $\tau$, 20000 synthetic realisations of the process~(\ref{ols}) were generated,
with $X:=t \in [-200, +1]$, in such a way that $Y^{data}$ displays a sudden change of regime at $\tau=-100$.
In the first half of the dataset  $[-200, -100]$, the data points are generated with $\beta = 0.3$.
In the second half of the dataset  $[-101, 0]$, the data points are generated with $\beta = 0.6$.
After the addition of random noise $\epsilon \sim \mathcal{N}(0, 1)$, each single resulting time-series 
was fitted for a fixed end time $\bar{t}_2$ = 1 while shrinking the left-most portion of the data ($t_1$) towards $t_2$, starting at $t_1=-200, -199, \ldots, t_2-3$. For the largest window with $t_1=-200$, there are $t_2-t_1+1=1-(-200)+1=202$ data points to fit. For the smallest window  with $t_1=t_2-3$, there are $t_2-t_1+1=4$ data points to fit.
For each window size $w$, the process of generating synthetic data and fitting the model was repeated 20000 times,
allowing us to obtain confidence intervals.

\begin{figure}[!h]
\begin{center}
\includegraphics[width=.5\textwidth]{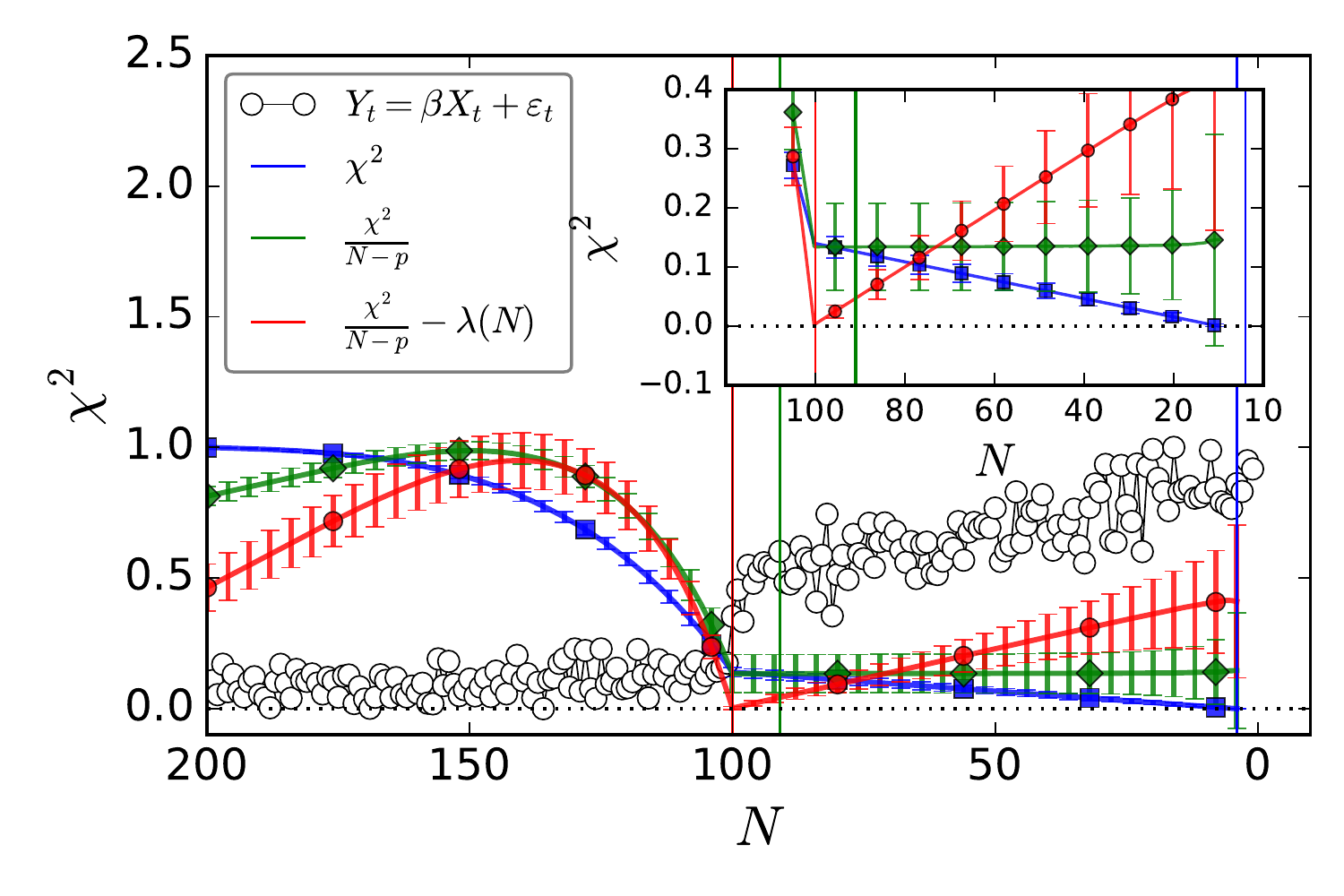}
\caption{{\bf Different goodness-of-fit measures applied to a shrinking-window linear regression problem (Eq. \ref{ols}) in order to diagnose the optimal calibration window length:} We simulated synthetic time-series with length N=200 (white circles) using 
expression (\ref{ols}) with a sudden change of regime at $t=-100$. We then fitted the same model (\ref{ols}) 
within shrinking windows (from left to right), i.e. for a fixed $t_2 = 1$, we shrink $t_1$ from $t_1$=-200 to $t_1$=-3 and show the values of $\chi^2(\bm{\Phi})$ (blue), $\chi^2_{np}(\bm{\Phi})$ (green) and $\chi^2_{\lambda}(\bm{\Phi})$ (red) metrics as a function of this shrinking estimation window. For each pair [$t_2:t_1$] (i.e. for each N), the process of generating synthetic data and fitting the model was repeated 20000 times (resulting on confidence bounds for each metric). For $t$=[-200:-100], $Y_t$ was simulated with $\beta = 0.3$ while from $t$ = [-100:1], $\beta=0.6$ was used. Without loss of generality, both the data and the cost functions had their values divided by their respectively maximum value in order to be bounded within the interval $[0,1]$. A Python script for generating the figure and performing all calculations can be found on Appendix.} \label{fig1}
\end{center}
\end{figure}

As depicted by Fig. (\ref{fig1}), the proposed methodology is able to correctly diagnose the optimal starting point := $\tau$
associated with the change of slope. 
While the $\chi^2(\Phi)$ metric monotonously decreases and the $\chi^2_{np}(\Phi)$ metric plateaus from $t=-100$ onwards, $\chi_{np}^2 - \lambda(t_2 - t_1)$ monotonously increases over the same interval, thus marking a clear minimum. The variance of the metric  $\chi^{2}_{\lambda}(\bm{\Phi})$ also increases over this interval.  Specifically, the metric   $\chi^2(\Phi)$ tends to favor the smallest windows and therefore overfitting is prone to develop and remain undetected.  The metric $\chi^{2}_{np}(\bm{\Phi})$ suggests $\tau \approx -90$ after 20000 simulations, which is 10\% away from the true value $\tau = 100$. Moreover, the dependence of $\chi^{2}_{np}(\bm{\Phi})$
as a function of $t_1$ is so flat for $t_1 \in [-100:-40]$ that any given value of $\tau$ within this period is statistically significant. For this simulation study, $\chi^{2}_{np}(\bm{\Phi})$ ranges for 0.134 to 0.135 for $t_1 \in [-100:-60]$, so as to be almost undistinguishable
over this interval of possible $\tau$ values. As we shall see later on, the performance of $\chi^{2}_{np}(\bm{\Phi})$ degrades further to resemble that of the $\chi^{2}(\bm{\Phi})$ metric when dealing with more complex nonlinear models such as the LPPLS model.
On the other hand, our proposed correction via the Lagrange regulariser $\lambda$ provides a simple and effective
method to identify the change of regime and the largest window size compatible with the second regime. The minimum is very pronounced and clear, which is not the case for $\chi^2_{np}(\bm{\Phi})$.

\section{Using the Lagrange regularisation method for Detecting the Beginning of Financial Bubbles}\label{sec4}

In the previous Section, we have proposed a novel goodness-of-fit metric for inferring the optimal beginning point or change point 
$\tau$ (for a fixed end point $\bar{t}_2$) in the calibration of a simple linear model.
The application of the Lagrange regulariser $\lambda$ allowed us to find the optimal window length $w^* = [\tau:t_2]$ for fitting the model by enabling the comparison of the goodness-of-fits across different $w$ values. We now extend the application of the methodology to a more complex non-linear model, which requires one to compare fits across different window sizes in order to diagnose bubble periods on financial instruments such as equity prices and price indexes.

\subsection{The LPPLS model}\label{sub_details}

The LPPLS (log-periodic power law singularity) model introduced by \cite{jls} provides a flexible set-up for diagnosing periods of price exuberance \citep{shiller:2000} on financial instruments. It highlights the role of herding behaviour, translating into positive feedbacks in the price dynamics
during the formation of bubbles. This is reflected in faster-than-exponential growth of the price of financial instruments.
Such explosive behavior is completely unsustainable and the bubbles usually ends with a crash or a progressive correction.
Here, we use the LPPLS model combined with the Lagrange regulariser $\lambda$
in order to detect the beginning of financial bubbles.

In the LPPLS model, the expectation of the logarithm of the price of an asset is written under the form
\begin{align}
fLPPL(\bm{\phi},t) &=A+B(f)+C_{1}(g)+C_{2}(h)
\label{fili},
\end{align}
where $\bm{\phi} = \{A,B,C_{1},C_{2},m,\omega,t_{c} \}$ is a $(1 \times 7)$ vector of parameters
we want to determine and
\begin{align}
f ~\equiv~ &(tc-t)^{m},\\
g ~\equiv~ &(t_{c}-t)^{m}\cos(\omega~\ln(t_{c}-t)),\\
h ~\equiv~ &(t_{c}-t)^{m}\sin(\omega~\ln(tc-t)).
\end{align}
Note that the power law singularity $(tc-t)^{m}$ embodies the faster-than-exponential growth.
Log-periodic oscillations represented by the cosine and sine of $\ln(t_{c}-t)$ model the long-term
volatility dynamics decorating the accelerating price. Expression (\ref{fili})
uses the formulation of \cite{FilimonovSornette2011_LPPL_calibration} in terms of 4 linear parameters $A,B,C_{1},C_{2}$
and 3 nonlinear parameter $m,\omega,t_{c}$.

Fitting Eq.~(\ref{fili}) to the log-price time-series amounts to search for the parameter set $\bm{\phi}^{*}$ that yields the smallest $N$-$dimensional$ distance between realisation and theory. Mathematically, using the $L^2$ norm, we form the following sum of squares of residuals 
\begin{align}
\resizebox{0.5\textwidth}{!}{$F(t_{c}, m,\omega,A,B,C_{1},C_{2})=\sum_{i=1}^{N} \Bigl{[} \ln[P(t_{i})] - A - B(f_{i})-C_{1}(g_{i})-C_{2}(h_{i}) \Bigr{]}^{2}$}\label{cost},
\end{align}
for $i = 1, \dots, N$. We proceed in two steps. 
First, enslaving the linear parameters $\{A,B,C_{1},C_{2}\}$ to the remaining nonlinear 
parameters $\bm{\phi} = \{t_{c}, m, \omega\}$, yields the cost function $\chi^{2}(\bm{\phi})$ 
\begin{align}
\chi^{2}(\bm{\phi}) &:= F_{1}(t_{c}, m,\omega)\\ &= \underset{\{A,B,C_{1},C_{2}\}} {\text{min}} F(t_{c}, m,\omega,A,B,C_{1},C_{2})\\ &= F(t_{c}, m, \omega, \widehat{A}, \widehat{B}, \widehat{C}_{1}, \widehat{C}_{2})~,
\label{F}
\end{align} 
where the hat symbol ~$\widehat{}$~ indicates estimated parameters.  
This is obtained by solving the optimization problem 
\begin{align}
\resizebox{0.5\textwidth}{!}{$\{\widehat{A}, \widehat{B}, \widehat{C}_{1}, \widehat{C}_{2} \}=arg\underset{\{A,B,C_{1},C_{2}\}} {\text{min}}~F(t_{c}, m,\omega. A,B,C_{1},C_{2}),$}\label{dacost}
\end{align}
which can be obtained analytically by solving the following system of equations,
\begin{equation}
\resizebox{0.5\textwidth}{!}{$\left[ \begin{array}{cccc}
N & \sum f_{i} &  \sum g_{i} &  \sum h_{i} \\
\sum f_{i} & \sum f_{i}^{2} & \sum f_{i}g_{i} & \sum f_{i}h_{i}\\
\sum g_{i} & \sum f_{i}g_{i} & \sum g_{i}^{2} & \sum g_{i}h_{i}\\
\sum h_{i} & \sum f_{i}h_{i} & \sum g_{i}h_{i} & \sum h_{i}^{2}\\
\end{array} \right]  \left[ \begin{array}{c}
\widehat{A} \\
\widehat{B} \\
\widehat{C}_{1} \\
\widehat{C}_{2} \\
\end{array} \right] =\left[ \begin{array}{c} 
\sum y_{i}  \\
\sum y_{i}f_{i} \\
\sum y_{i}g_{i} \\
\sum y_{i}h_{i} \\
\end{array} \right].$}
\end{equation}

Second, we solve the nonlinear optimisation problem involving the remaining nonlinear 
parameters $m,\omega,t_{c}$:
\begin{align}
\{\widehat{t}_{c}, \widehat{m}, \widehat{\omega}\}&= arg \underset{\{t_{c}, m,\omega\}}{\text{min}} ~ F_{1}(t_{c},  m, \omega).\label{costtt}
\end{align} 
The model is calibrated on the data using the Ordinary Least Squares method, providing estimations of all parameters $t_c$, $\omega$, $m$, $A$, $B$, $C_1$, $C_2$ in a given time window of analysis. 

For each fixed data point $t_2$ (corresponding to a fictitious ``present'' up to which the data is recorded), we fit 
the price time series in shrinking windows $(t_1,t_2)$ of length $dt:=t_2-t_1$ decreasing from 1600 trading days to 30 trading days. We shift the start date $t_1$ in steps of 3 trading days, thus giving us 514 windows to analyse
for each $t_2$. In order to minimise calibration problems and 
address the sloppiness of the model with respect to some of its parameters
(and in particular $t_c$), we use a number of filters to select the solutions. For further information about the sloppiness of the LPPLS model, we refer to \citep{BreeChallet2010,Sornette_etal2015_Shanghai, DemosSornette2015, lppls:modLikelihood2016}. 
The filters used here are $\{(0.1 < m < 0.9), (6 < \omega < 13), (t_2-[t2-t1] < t_c < t_2+[t2-t1])\}$, so that
only those calibrations that meet these conditions are considered valid and the others are discarded.
These filters derive from the empirical evidence
gathered in investigations of previous bubbles \citep{ZhouSornette2003,QunQunzhididier15, Sornette_etal2015_Shanghai}.

Previous calibrations of the JLS model
have further shown the value of additional constraints imposed on the 
nonlinear parameters in order to remove spurious calibrations (false positive identification of bubbles)
 \citep{DemosSornette2015,bree,GeraskinFanta2011}. For our purposes, we do not consider them here.

\subsection{Empirical analysis}\label{sec5}

We apply our novel goodness-of-fit metric to the problem of finding the beginning times of financial bubbles,
defined as the optimal starting time $t_1$ obtained by endogenising $t_1$ and calibrating it.
We first illustrate and test the method on synthetic time series and then apply it to
real-world financial bubbles. A Python implementation of the algorithm is provided in the appendix. 

\subsubsection{Construction of synthetic LPPLS bubbles \label{rwrtbgqq}}

To gain insight about the application of our proposed calibration methodology on a controlled framework and thus 
establish a solid background to our empirical analysis, we generate synthetic price time series
that mimic the salient properties of financial bubbles, namely, a power law-like acceleration decorated by oscillations.
The synthetic price time series are obtained by using formula (\ref{fili}) with parameters given by the best LPPLS fit within the window $w \in$ $[t_{1} = 1 ~Jan.~1981$:~$t_{2} = 30 ~Aug.~1987]$ of the bubble that ended with 
the Black Monday 19 Oct. 1987 crash. 
These parameters are m = 0.44, $\omega$=6.5, $C_{1}$ = -0.0001, $C_{2}$=0.0005, $A$=1.8259, $B$= -0.0094, $t_{c}$ = 1194 (corresponding to 1987/11/14),
where days are counted since an origin put at  $t_{1} = Jan.~1981$.  To the deterministic component describing the expected log-price given by expression (\ref{fili}) and
denoted by $fLPPLS(\bm{\phi},t)$,
we add a stochastic element to obtain the synthetic price time series
\begin{align}
\ln[P(t)] = fLPPLS(\bm{\phi},t) + \sigma \epsilon(t),
\label{thjryuj4ew}
\end{align}
where $\epsilon(t)$  $\sim \mathcal{N}(0,\sigma^0)$ noise, $\sigma^0 = 0.03$ and $t = [1,\dots ,N=1100]$. 

\begin{figure}[!h]
\begin{center}
\includegraphics[width=.49\textwidth]{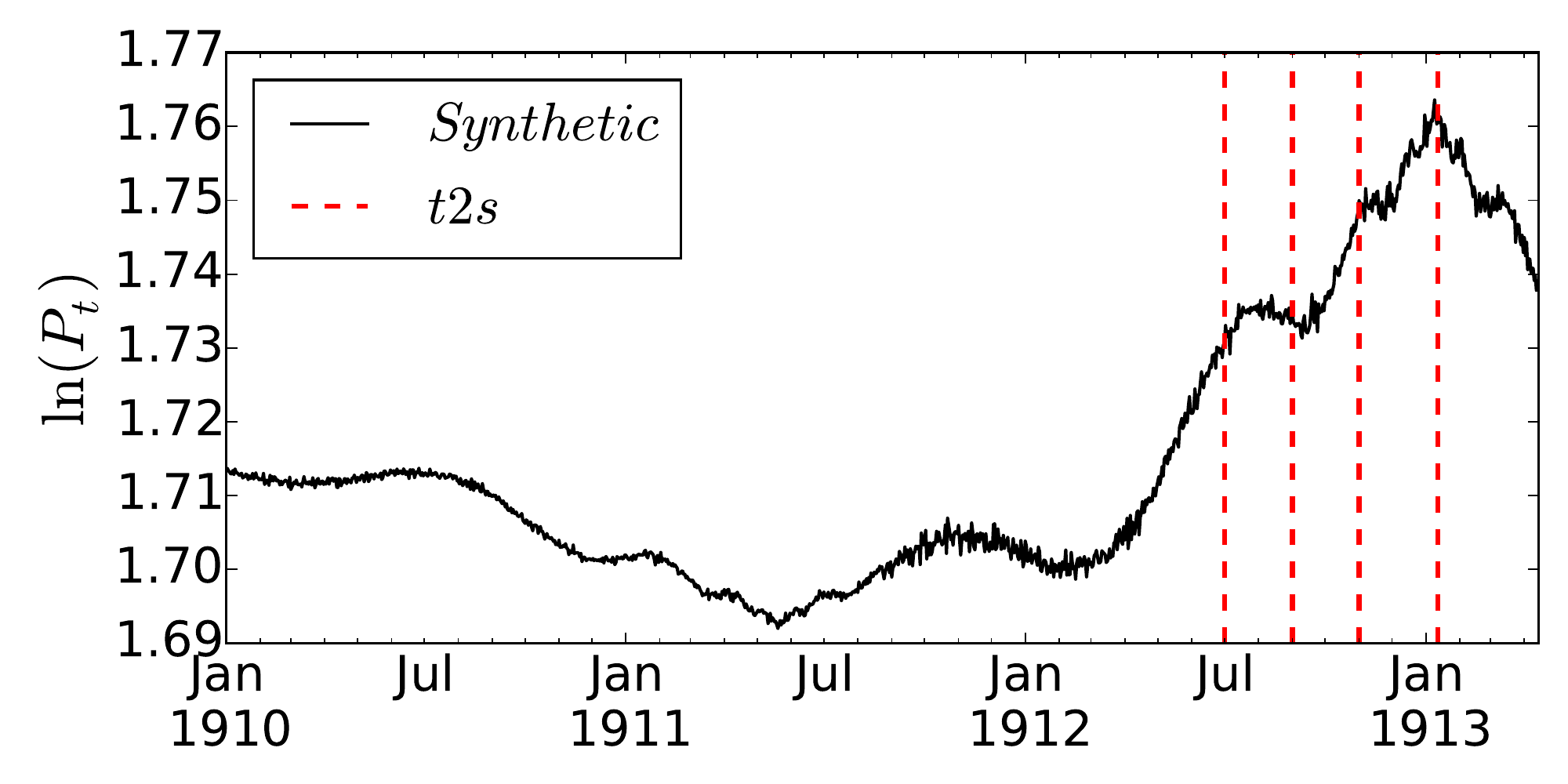}
\includegraphics[width=.49\textwidth]{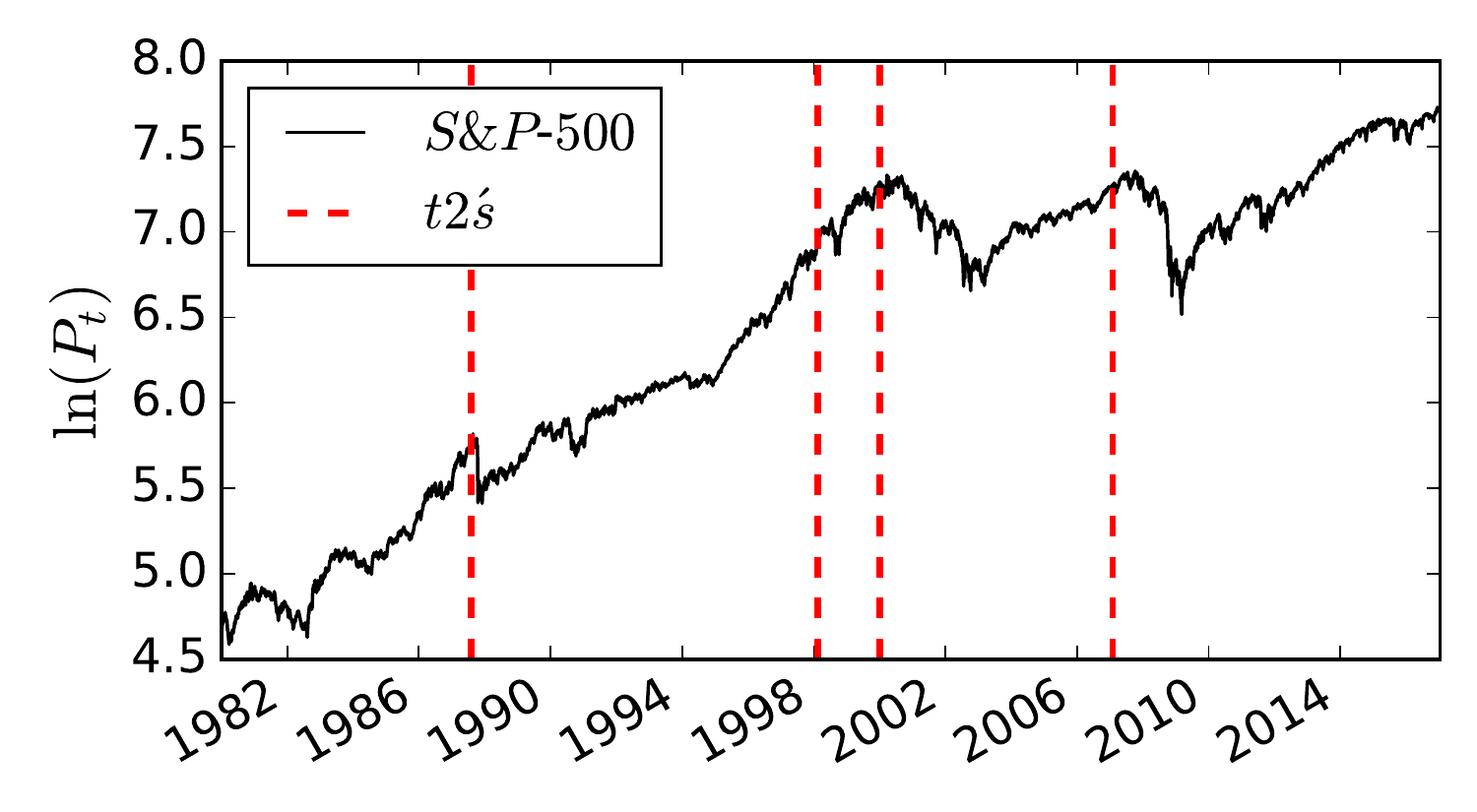}
\includegraphics[width=.49\textwidth]{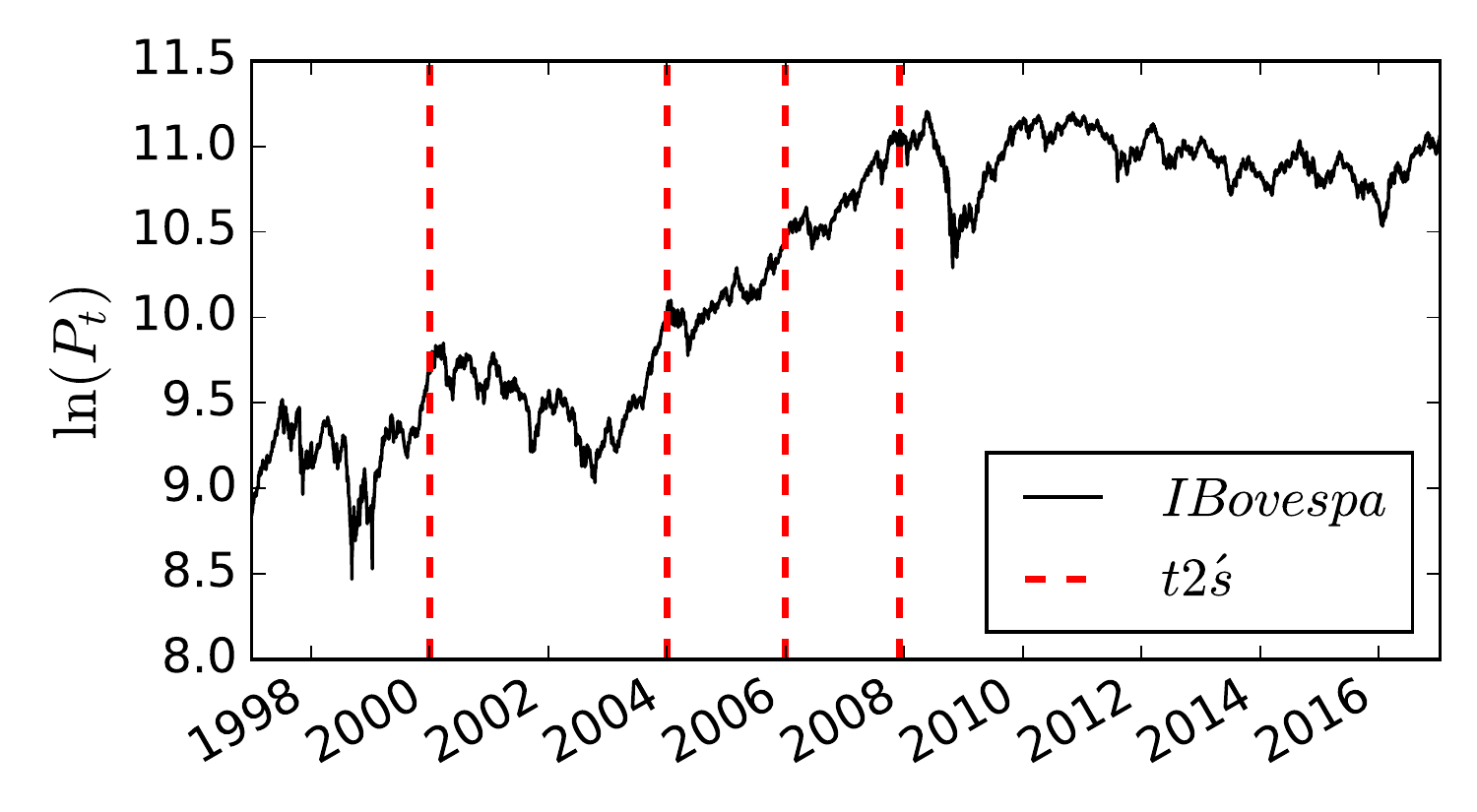}
\includegraphics[width=.49\textwidth]{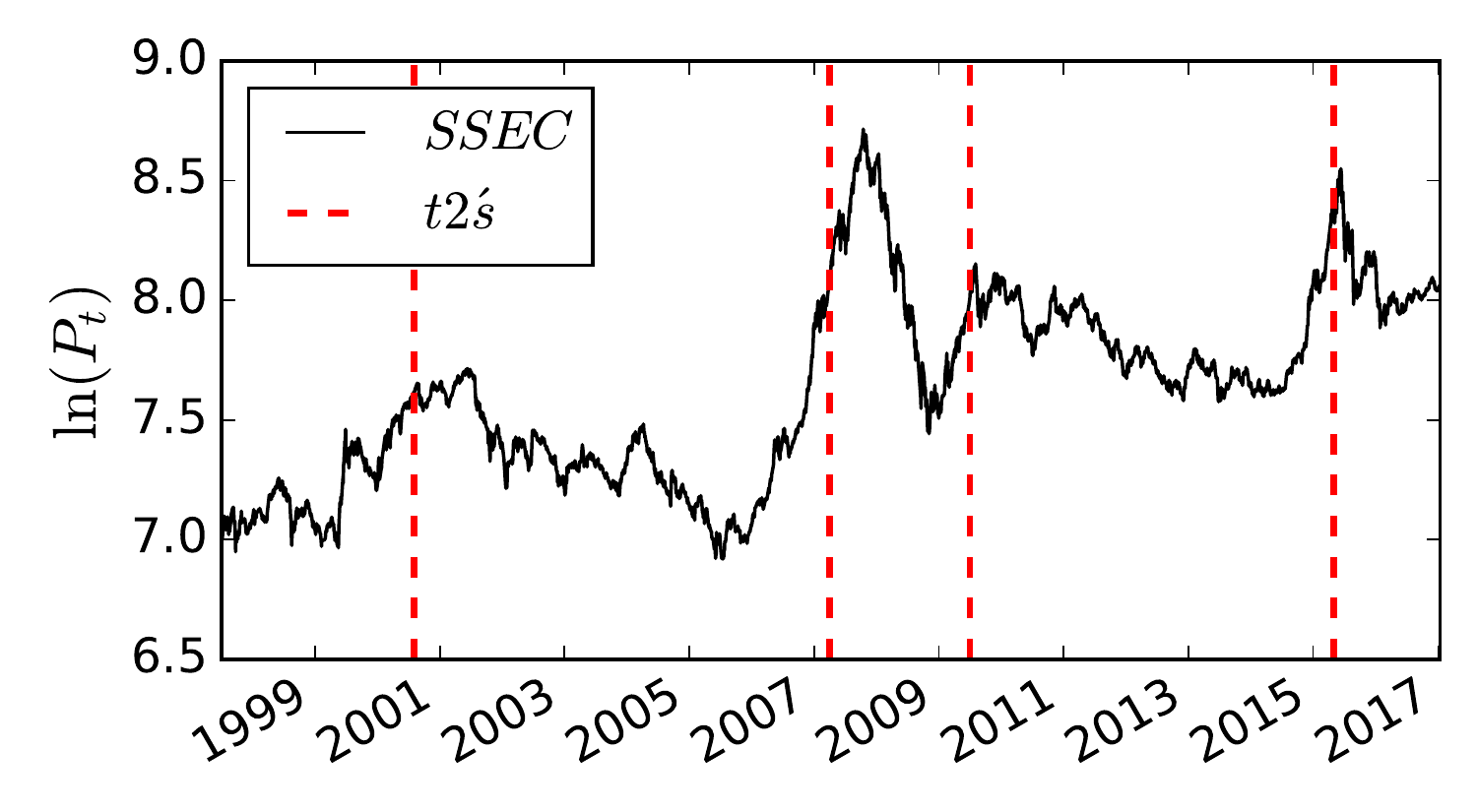}
\caption{{\bf Synthetic and real-world Time-series used in this study for measuring the performance of different goodness-of-fit metrics at different $t_2$'s (red lines):} Synthetic time-series and Indexes $S\&P$-$500$, $IBovespa$ and $SSEC$ with $t_2's = \{1912.07.01; 1912.10.01; 1912.11.15; 1913.01.01\}$,  $t_2's = \{1987.07.15;\, 1997.06.01;\, 2000.01.01;\, 2007.06.01\}$, $t_2's = \{2000.01.01; 2004.01.01; 2006.01.01; 2007.12.01\}$ and $t_2's = \{2000.08.01; 2007.05.01; 2009.07.01; 2015.05.01\}$ respectively (red dashed vertical lines).}\label{figX}
\end{center}
\end{figure}

To create a price time series with a well-defined transition point corresponding to the beginning of a bubble,
we take the first 500 points generated with expression (\ref{thjryuj4ew}) and mirror them
via a $t \to t_1 -t$ reflection across the time $t_{1} = 1 ~Jan.~1981$.
We concatenate this reflected sequence of 500 prices to the 1100 prices obtained with (\ref{thjryuj4ew})
for $t \geq t_1$, so that the true transition point corresponding to the start of the bubble described by the LPPLS pattern
is $t_{1} = 1 ~Jan.~1981$. The black stochastic line on the top of figure (\ref{figX}) represent this union of the two time-series. This union constitutes the whole synthetic 
time series on which we are going to apply our Lagrange regularisation of  $\chi^{2}_{\lambda}(\bm{\Phi})$
in order to attempt recovering the true start time, denoted by the hypothetical time $t_{1} = 1 ~Jul.~1911$.

For each synthetic bubble price time series, we thus calibrated it with 
Eq. (\ref{fili}) by minimizing expression (\ref{1}) in windows $w = [t_1, t_2]$, varying $t_2$ from 1912/07/01 to $t_2=1913/01/01$, with $t_1$ scanned from $t_{1} = Jan.~1910$ up to 30 business days before $t_{2}$, i.e. up to $t_{1,max} = t_{2}-30$ for each fixed $t_2$. The goal is to determine whether the transition point $\tau$ we determine is close (or even equal to)
the true hypothetical value $t_{1} = 01 ~Jul.~1911$ for different maturation times $t_2$ of the bubble. The number of degrees of freedom used for this exercise as well as for the real-world time series is $p=8$, which includes the 7 parameters
of the LPPLS model augmented by the extra parameter $t_1$.

\subsubsection{Real-world data: analysing bubble periods of different financial Indices \label{ssss}}

The real-world data sets used consists on bubble periods that have occurred on the following major Indexes: $S\&P$-$500$\footnote{$t_2$'s = $\{$1987.07.15;\, 1997.06.01;\, 2000.01.01;\, 2007.06.01$\}$}, IBovespa\footnote{$t_2$'s = $\{$2000.01.01;\, 2004.01.01;\, 2006.01.01;\, 2007.12.01$\}$} and $SSEC$\footnote{$t_2$'s = $\{$2000.08.01;\, 2007.05.01;\,  2009.07.01;\, 2015.05.01$\}$}. For each data set and for each fixed pseudo present time $t_2$ depicted by red vertical dashed lines on Fig.~(\ref{figX}), our search for  the bubble beginning time $\tau$ consists in fitting the LPPLS model using a shrinking estimation window $w$ with $t_1 = [t_2-30:t_2-1600]$ with incremental step-size of 3 business days. This yields a total of 514 fits per $t_2$.

\subsubsection{Analysis \label{analysis}}

\begin{figure}[!tp]
\begin{center}
\includegraphics[width=.4\textwidth]{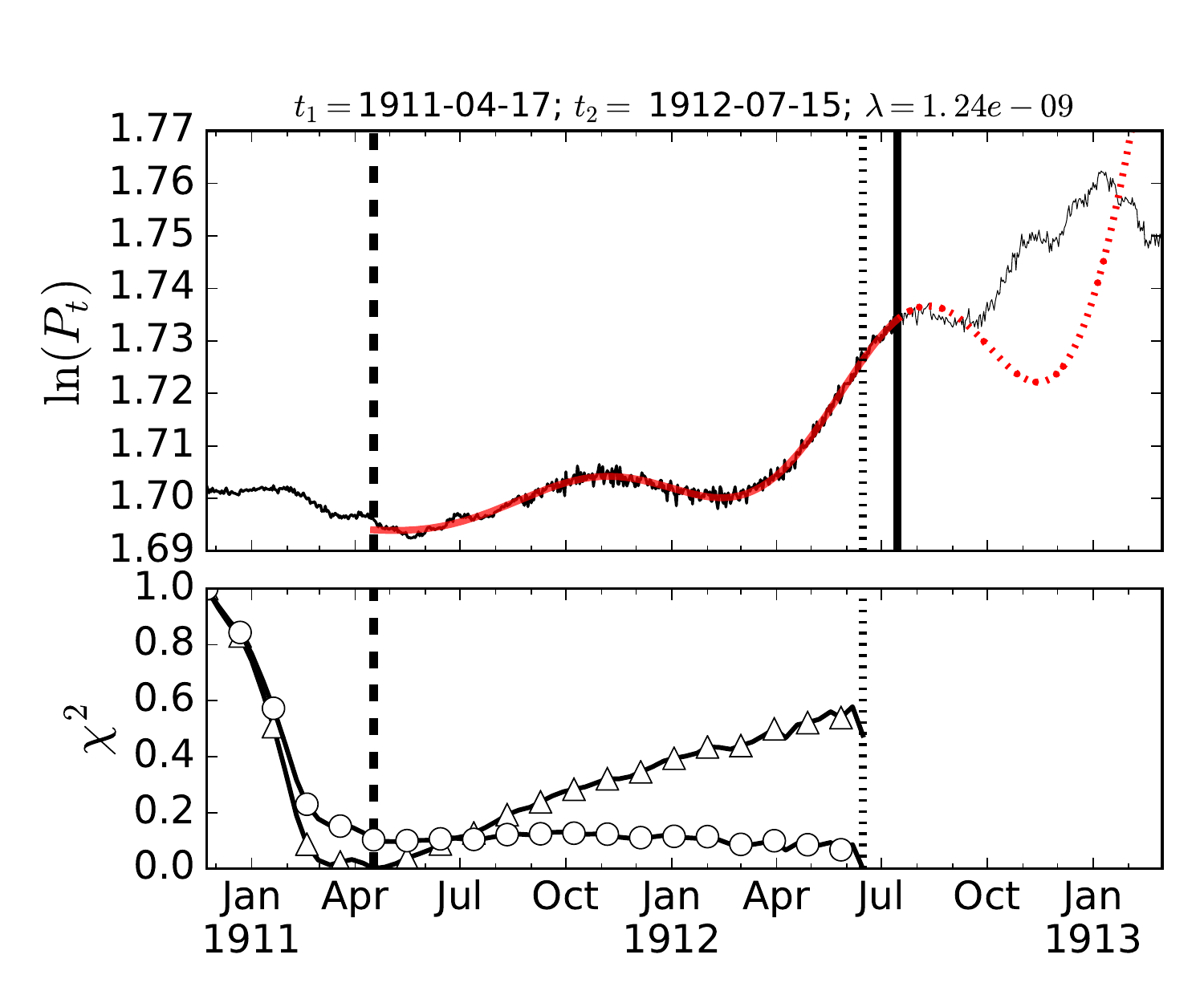}
\includegraphics[width=.4\textwidth]{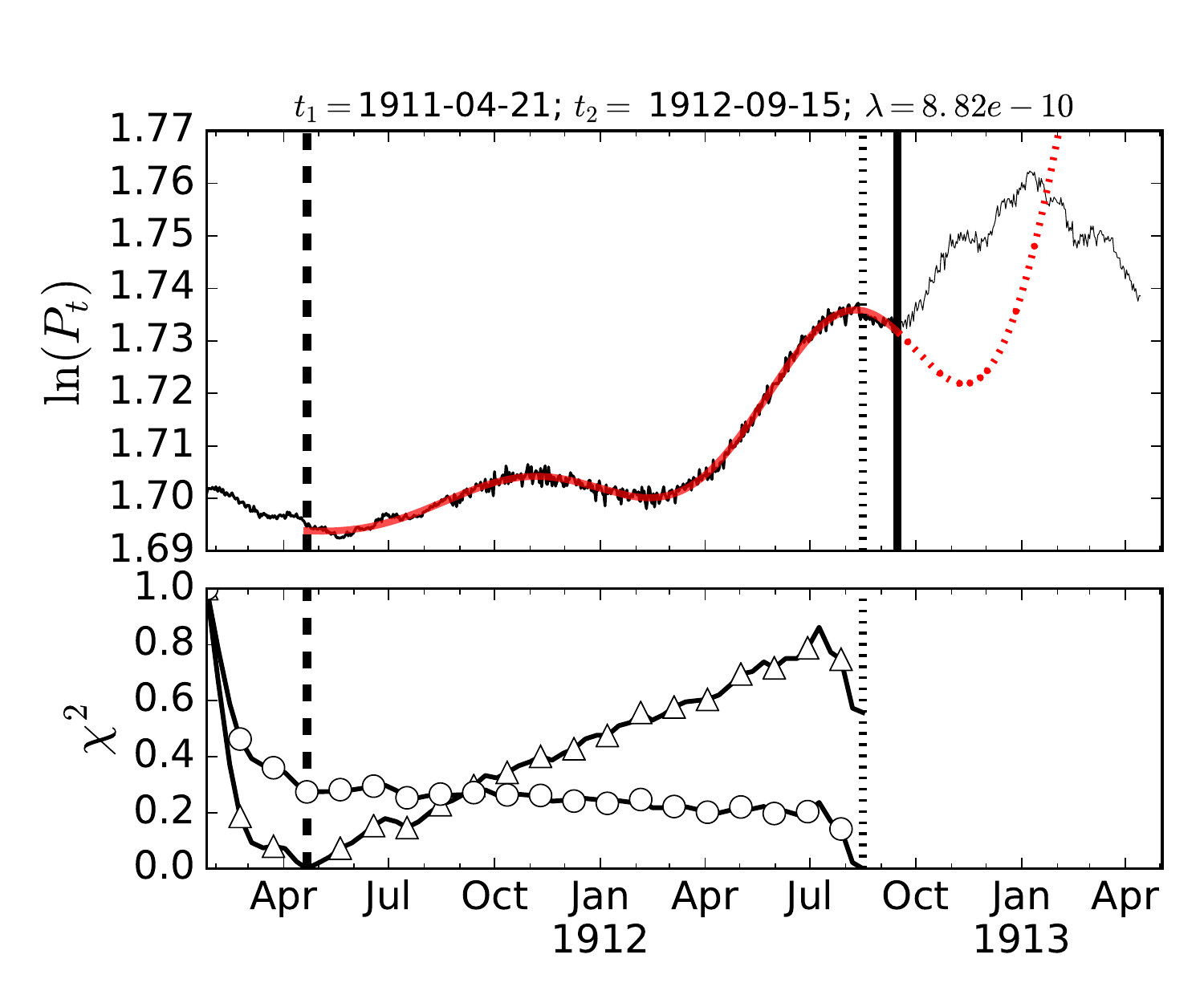}
\includegraphics[width=.4\textwidth]{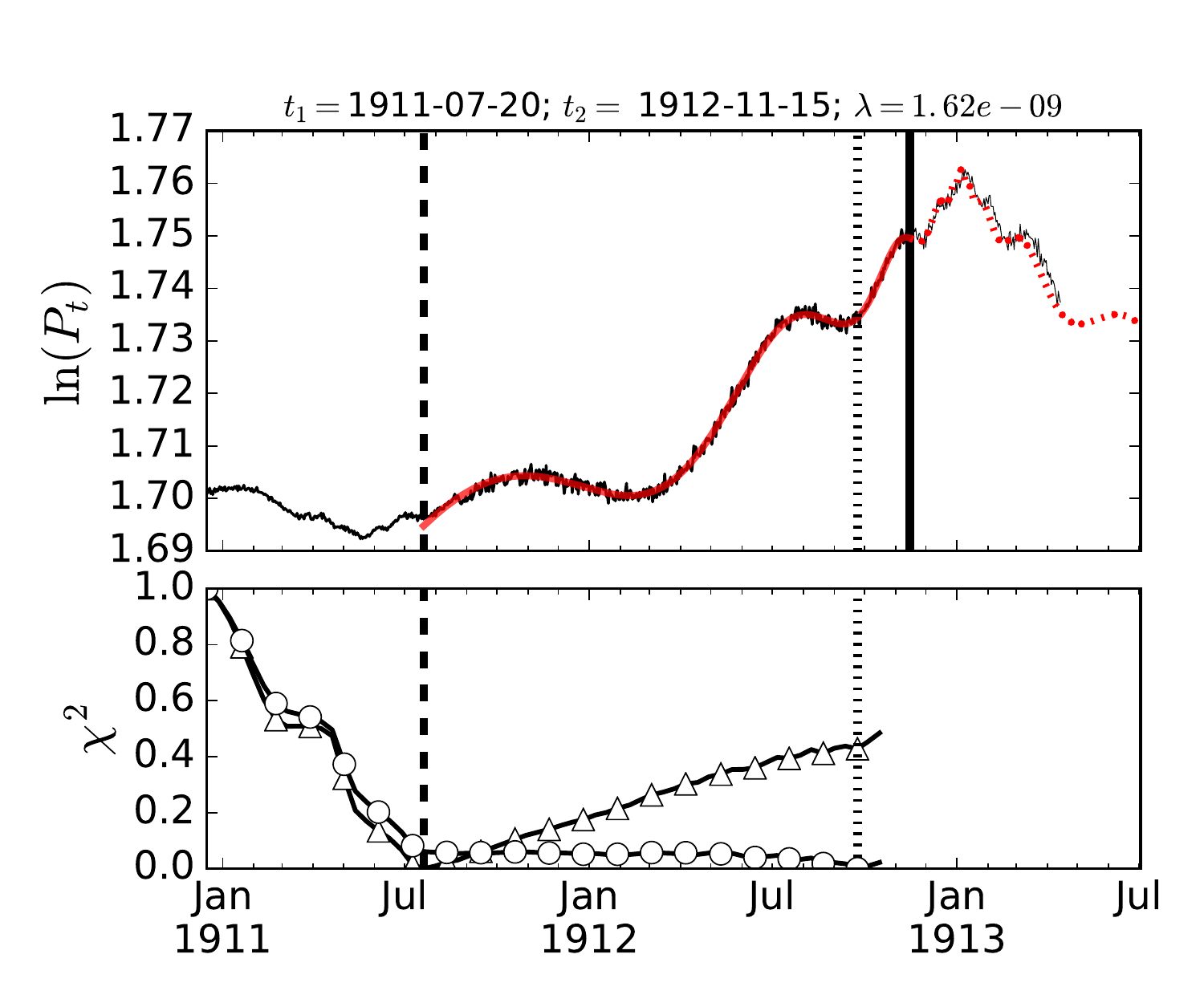}
\includegraphics[width=.4\textwidth]{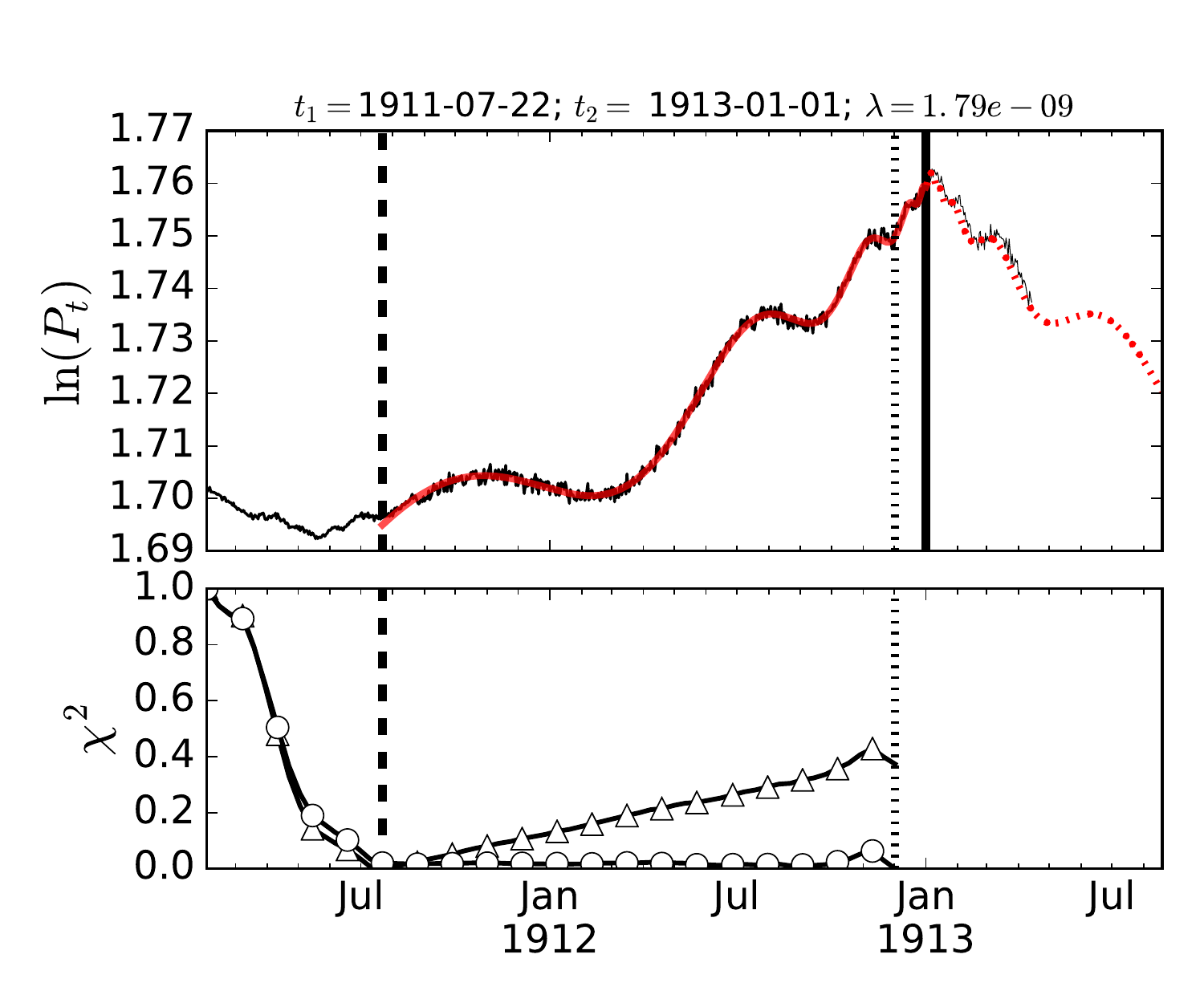}
\caption{{\bf Diagnosing the beginning of financial bubbles by comparing two goodness-of-fit metrics $\chi^{2}_{np}(\bm{\Phi})$ vs.
$\chi^{2}_{\lambda}(\bm{\Phi})$ using the LPPLS model on Synthetic Time-Series:} $\chi^{2}_{np}(\bm{\Phi})$ is depicted by blank circles in the lower plot while our proposed metric is depicted by blank triangles. The dashed black vertical lines denotes the minimum of each goodness of fit metric and therefore represents the optimal $\tau \equiv t_1$ for $\chi^{2}_{np}(\bm{\Phi})$ and $\chi^{2}_{\lambda}(\bm{\Phi})$. For a fixed $t_2$, the log-price time-series of the Index was fitted using a shrinking window from $t_1 = [t_2 - 30: t_2-1600]$ sampled every 3 days. For a fixed $t_2$ and $t_1$, we display the resulting fit of the LPPLS model (red line) obtained with the parameters solving Eq. (\ref{costtt}).}\label{fig3}
\end{center}
\end{figure}

Let us start with the analysis of the synthetic time-series\footnote{$t_2$'s = $\{1912.07.01;~1912.10.01;~1912.11.15;~1913.01.01\}$} depicted in Fig.~(\ref{fig3}). 
For the earliest $t_2$ = 1912/07/01, our proposed 
goodness-of-fit scheme is already capable of roughly diagnosing correctly the bubble beginning time,
finding the optimal $\tau$ to be $\approx May~ 1911$.
In contrast, the competing metric ($\chi^2_{np}(\bm{\Phi})$) is degenerate as $t_1 \rightarrow t_2$
and is thus blind to the beginning of the bubble.
For $t_2$ closer to the end of the bubble, $\chi^2_{np}(\bm{\Phi})$ continues to deliver very small optimal windows,
proposing the incorrect conclusion that the bubble has started very recently (i..e close to the pseudo 
present time $t_2$). This is a signature of strong overfitting, which is quantified via $\lambda$ and depicted in the title of the figure alongside with the bubble beginning time and $t_2$. The 
Lagrange regularisation of the $\chi^2_{np}(\bm{\Phi})$ locks into the true value of $\tau \approx Jul.1911$ as $t_2 \rightarrow t_c$, i.e., as $t_2$ moves closer and closer to January 1913 and the LPPLS signal becomes stronger.

\begin{figure}[!tp]
\begin{center}
\includegraphics[width=.4\textwidth]{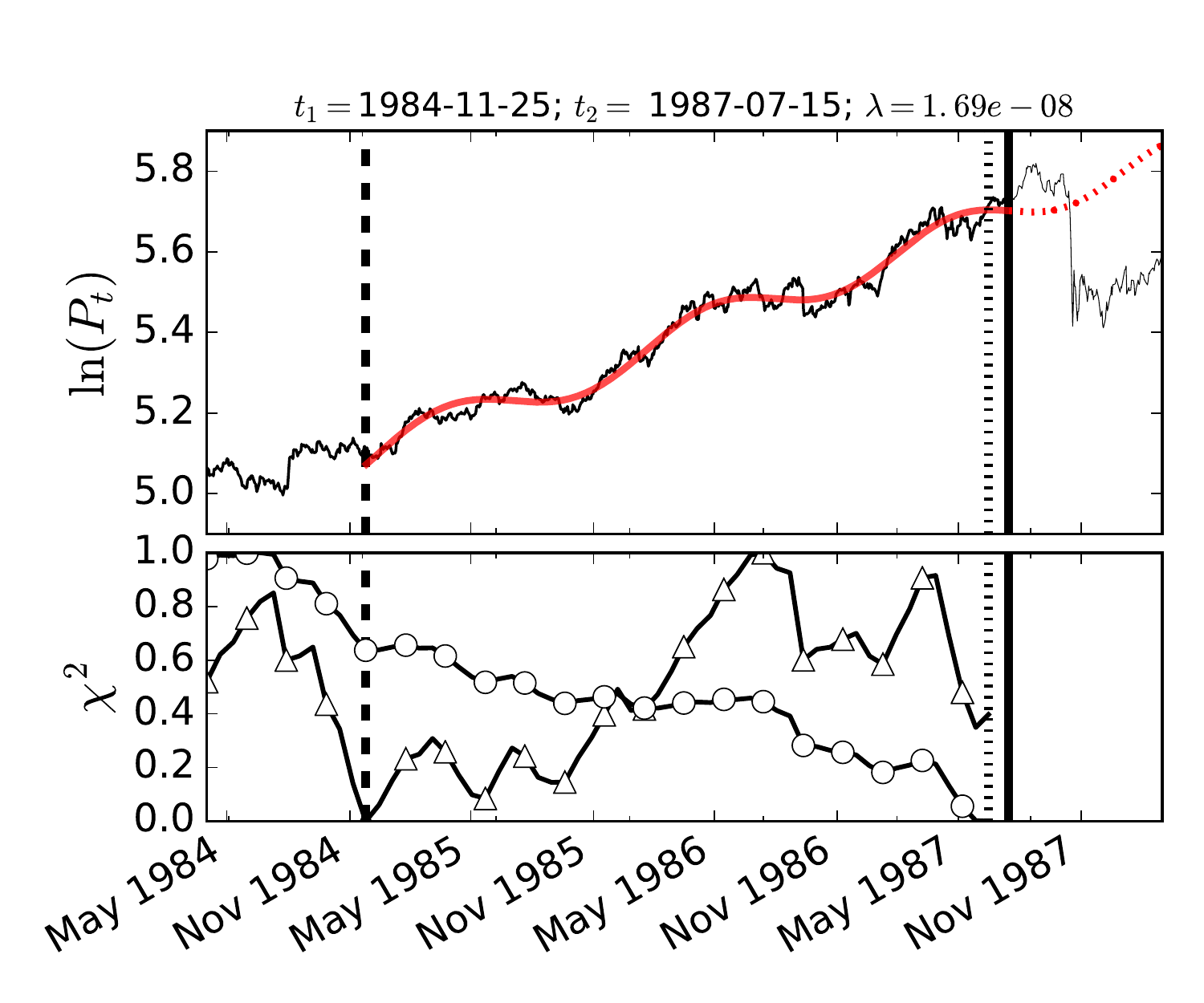}
\includegraphics[width=.4\textwidth]{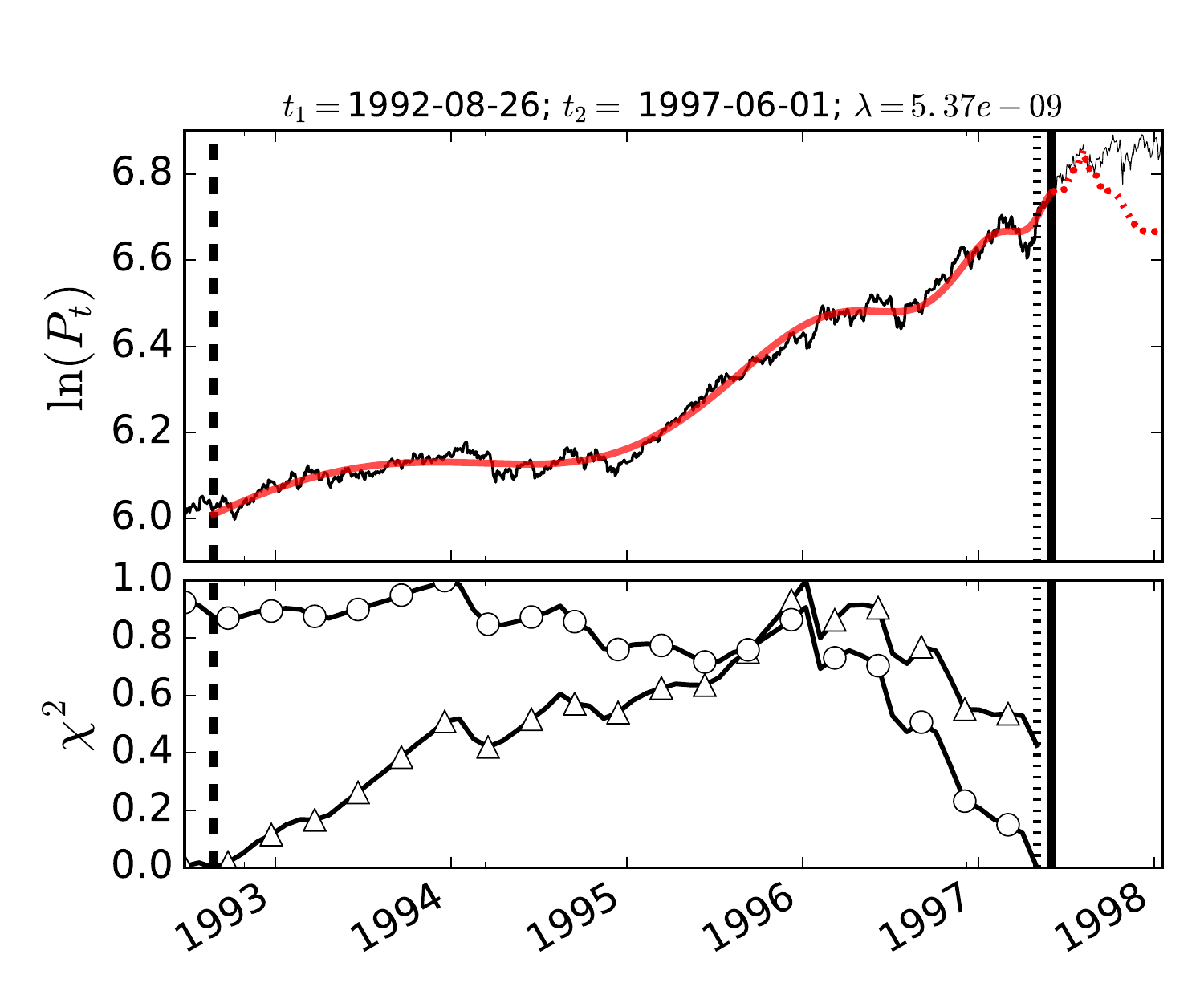}\\
\includegraphics[width=.4\textwidth]{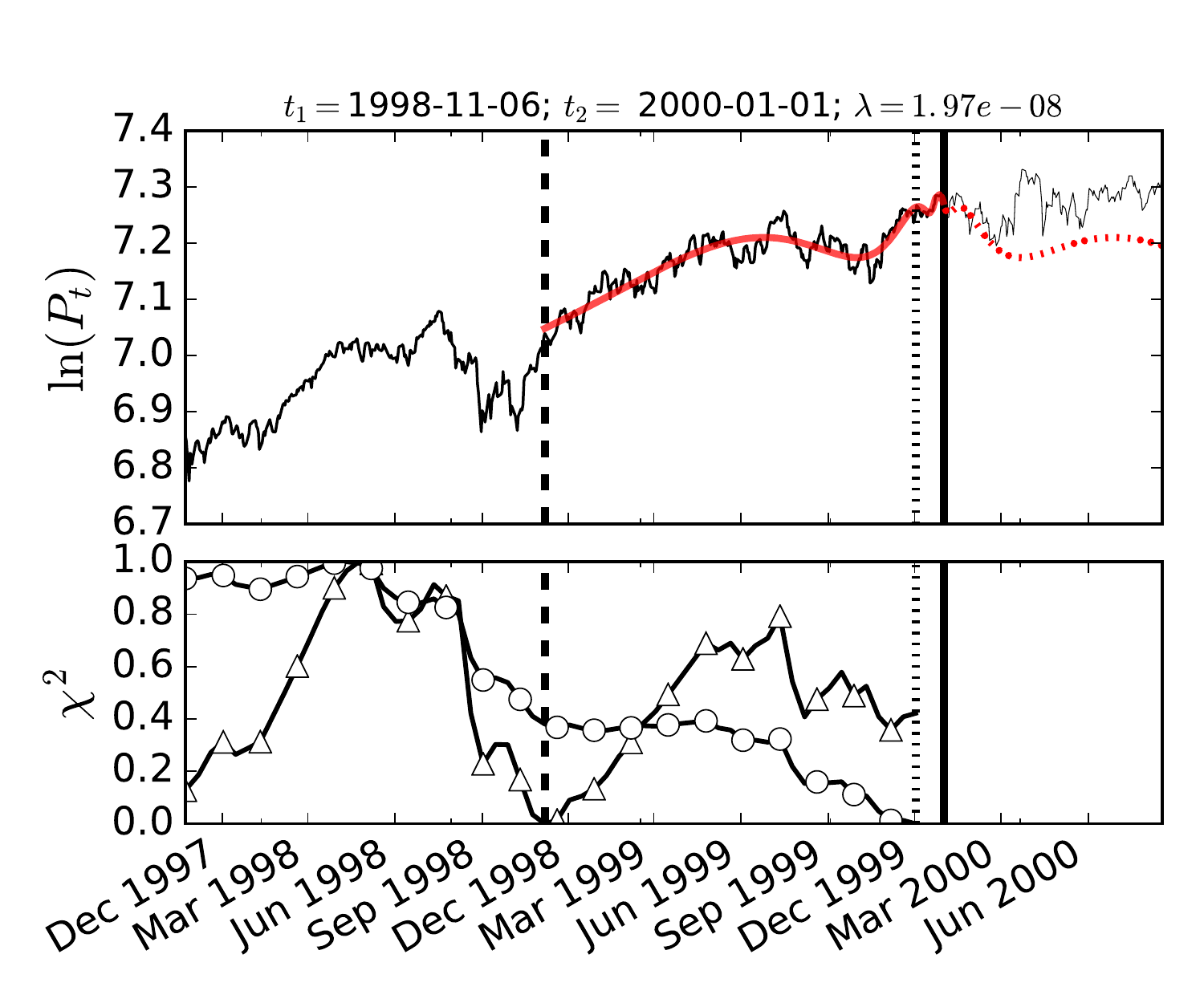}
\includegraphics[width=.4\textwidth]{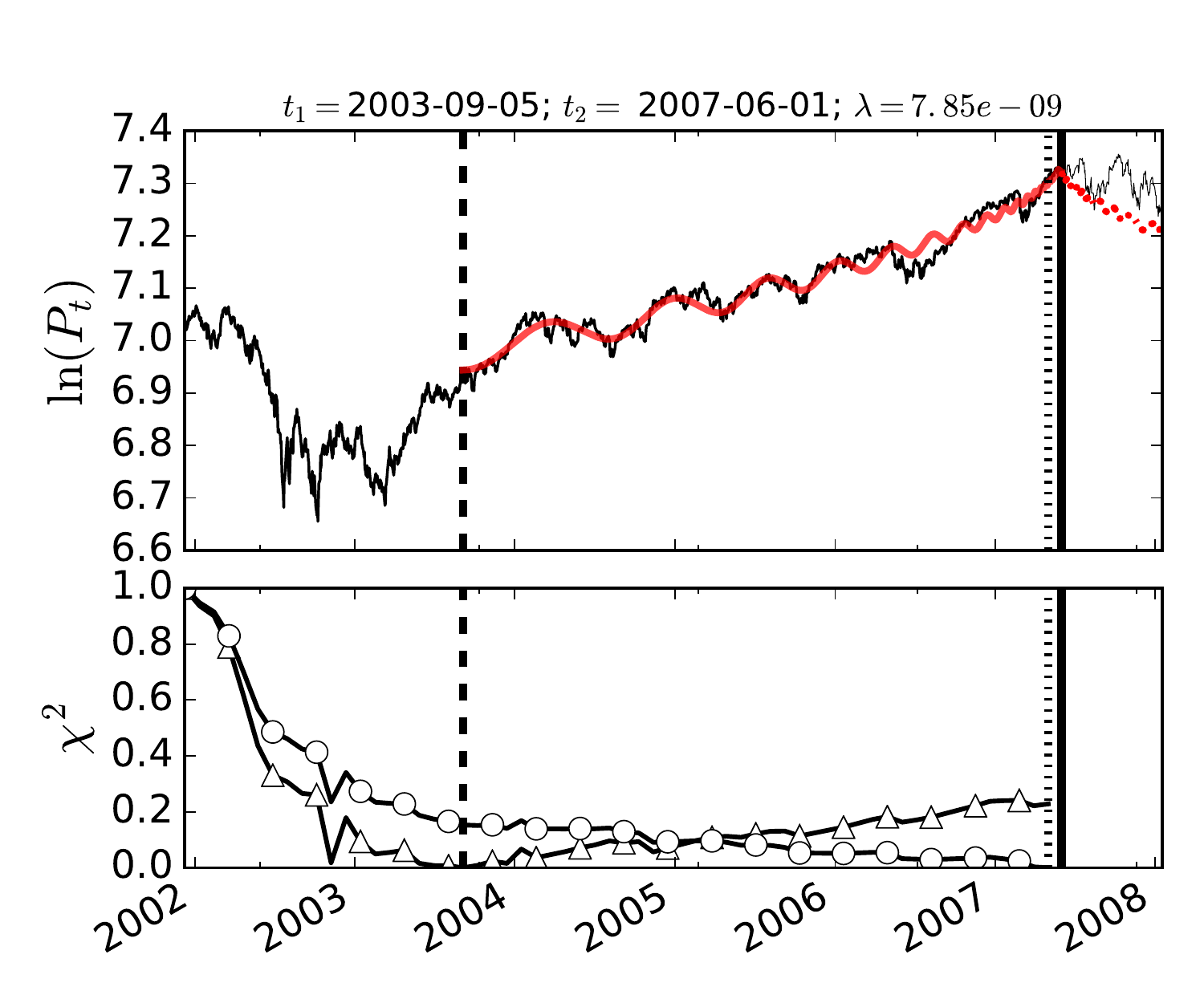}
\caption{{\bf Same as figure \ref{fig3} for the US $S\&P$-$500$ Index.} }\label{fig4}
\end{center}
\end{figure}

We now switch to the real-world time-series. For the $S\&P$-$500$ Index, see Fig.~(\ref{fig4}), the results obtained are even more pronounced. While again $\chi^2_{np}(\bm{\Phi})$ is unable to diagnose the optimal starting date of a faster than exponential log-price growth $\tau \equiv t_1$, the Lagrange regularisation of the $\chi^2_{np}(\bm{\Phi})$
 depicted by blank triangles in the lower box of the figure is capable of overcoming the tendency of the model to overfit data as $t_1 \rightarrow t_2$. Specifically, the method diagnoses the start of the Black-Monday bubble at $t_1 \approx March~1984$ and the beginning of the Sub-Prime bubble at $\approx$ Aug.~2003 in accordance with \citep{ZhouSor05}.

We also picked two pseudo present times $t_2's$ at random in order to 
check how consistent are the results. To our delight, the method is found capable of capturing the different time-scales present of bubble formation in an endogenous manner. For $t_2 = 1997.06.01$, the method
suggests the presence of a bubble that nucleated more than five years earlier. This recovers the bubble and change
of regime in September 1992, documented in Chapter 9 of \citep{sornette2003} as a ``false alarm''
in terms of being followed by a crash. Nevertheless, it was a genuine change of regime as the market stopped its ascent
and plateaued for the three following months. 
For $t_2$ = 2000.01.01, $\chi^{2}_{\lambda}(\bm{\Phi})$
 diagnoses a bubble with a shorter duration, which started
in November 1998. The starting time is coherent with the recovery after the so-called Russian crisis of August-September 1998
when the US stock markets dropped by about 20\%. And this bubble is nothing but the echo in the S\&P500
of the huge dotcom bubble that crashed in March-April 2000.
More generally, scanning $t_2$ and different intervals for $t_1$, the Lagrange regularisation of the 
$\chi^{2}_{np}(\bm{\Phi})$ can
endogenously identify a hierarchy of bubbles of different time-scales, reflecting their multi-scale structure
\citep{sornette2003,lppls:modLikelihood2016}. 

\begin{figure}[!h]
\begin{center}
\includegraphics[width=.4\textwidth]{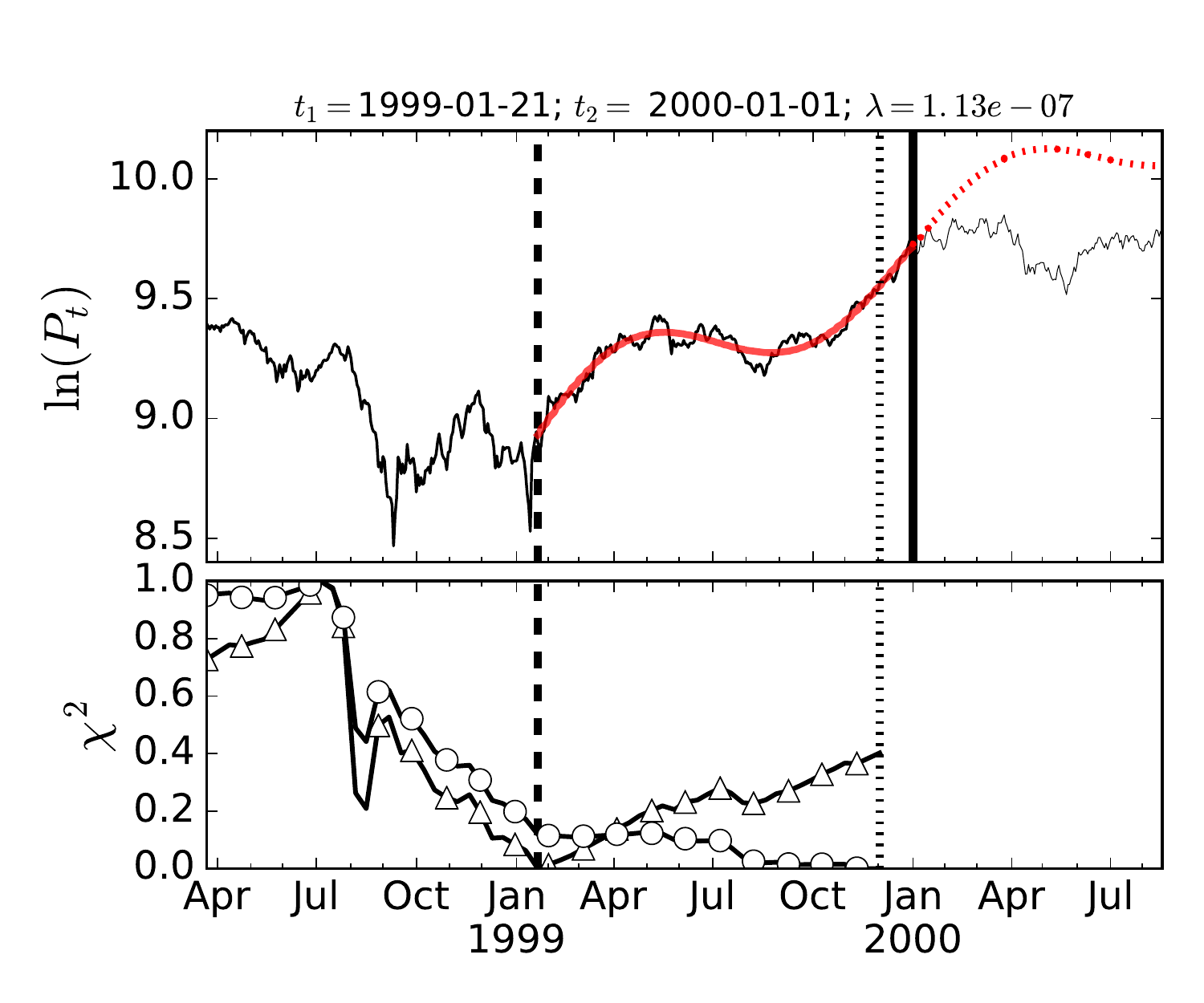}
\includegraphics[width=.4\textwidth]{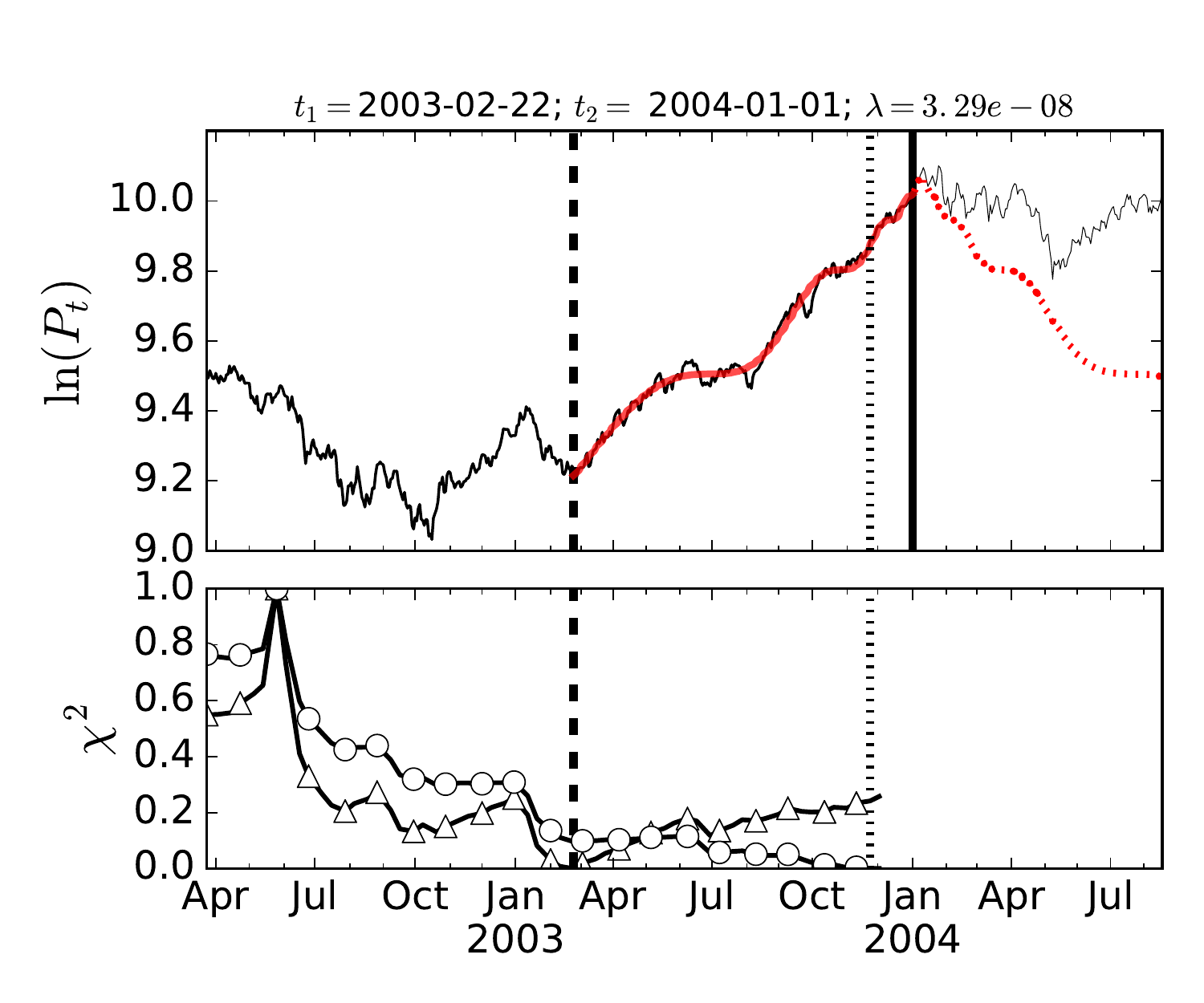}\\
\includegraphics[width=.4\textwidth]{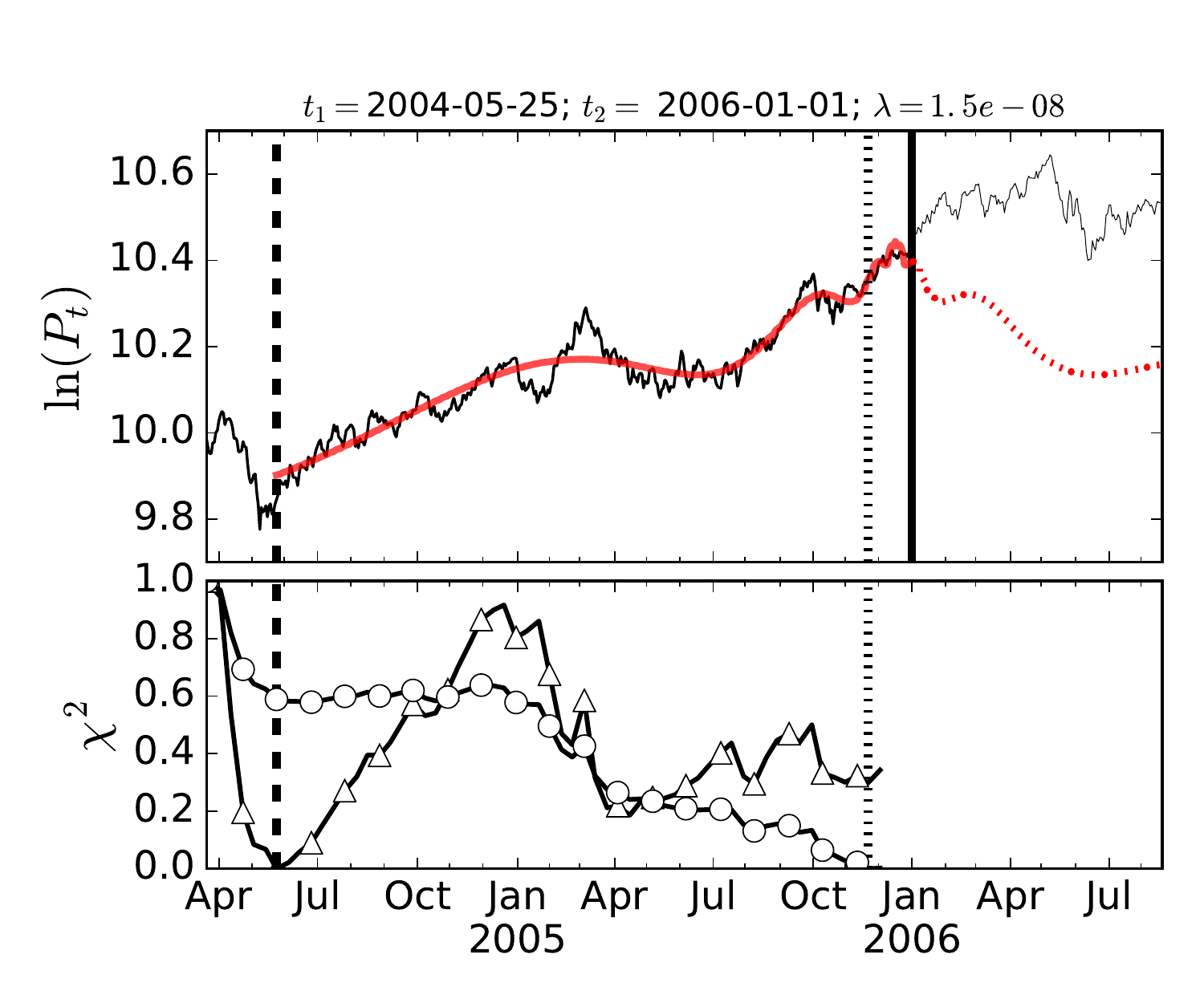}
\includegraphics[width=.4\textwidth]{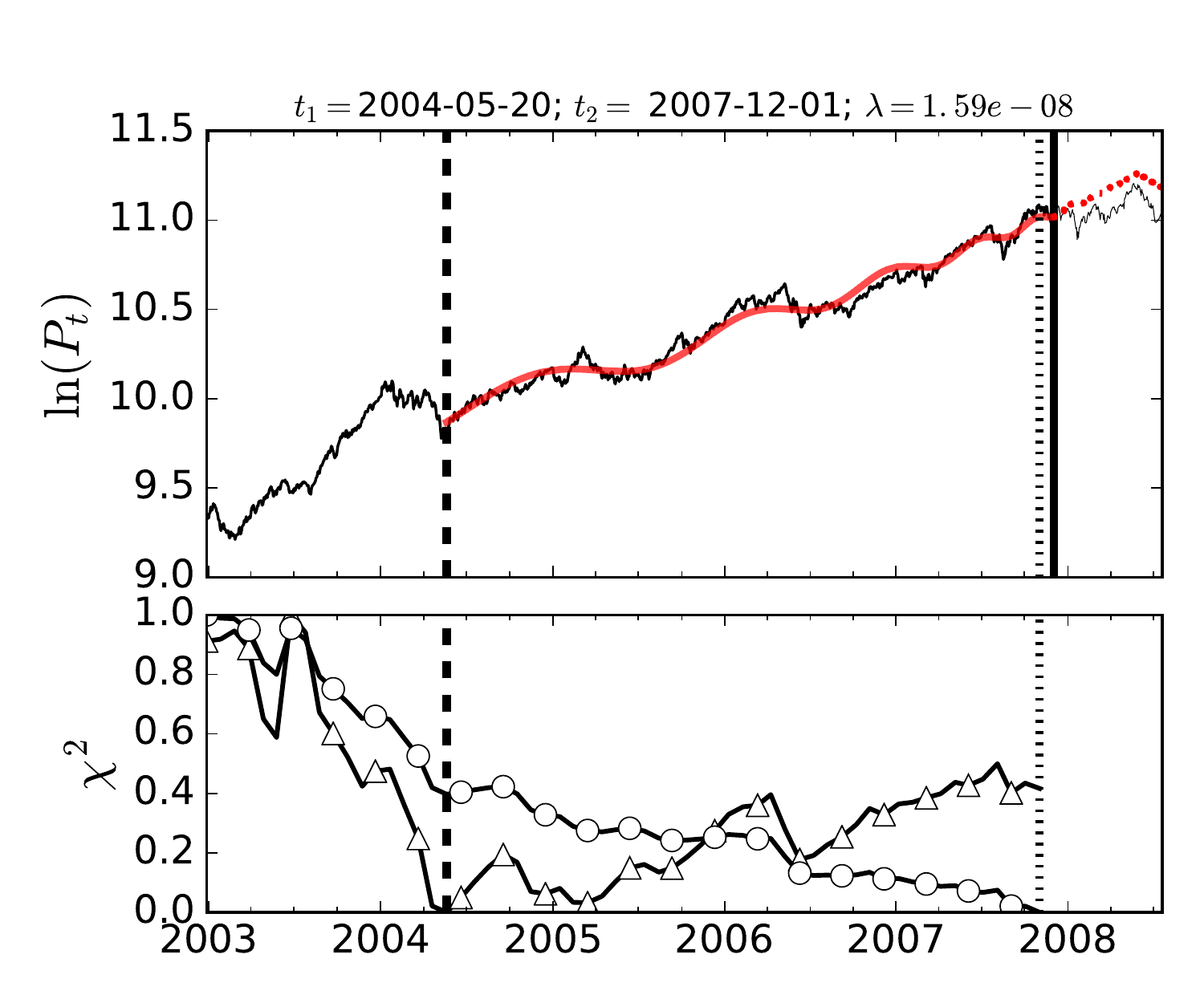}
\caption{{\bf Same as figure \ref{fig3} for the Brazilian $IBovespa$ index.}}\label{fig5}
\end{center}
\end{figure}

\begin{figure}[!h]
\begin{center}
\includegraphics[width=.4\textwidth]{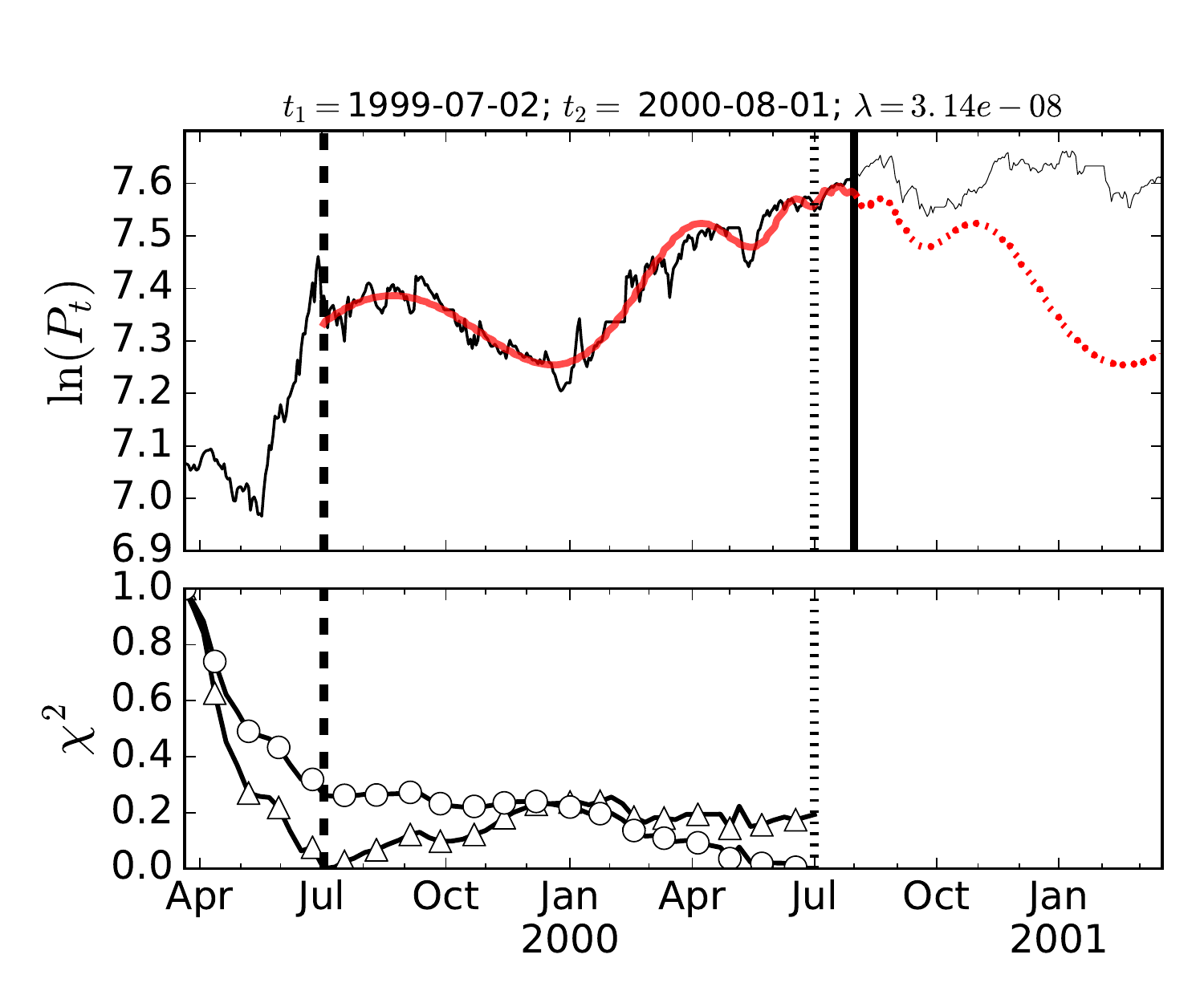}
\includegraphics[width=.4\textwidth]{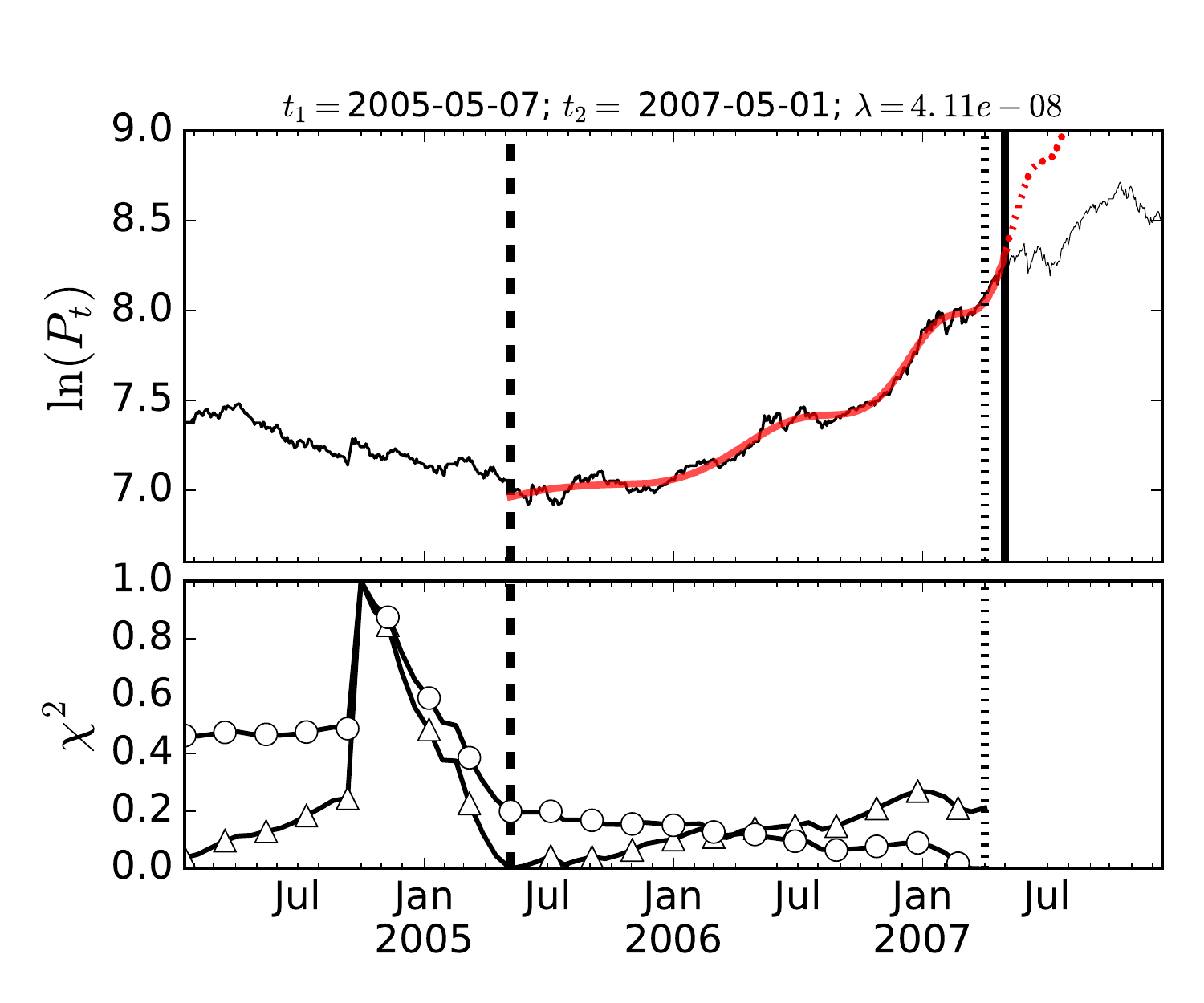}\\
\includegraphics[width=.4\textwidth]{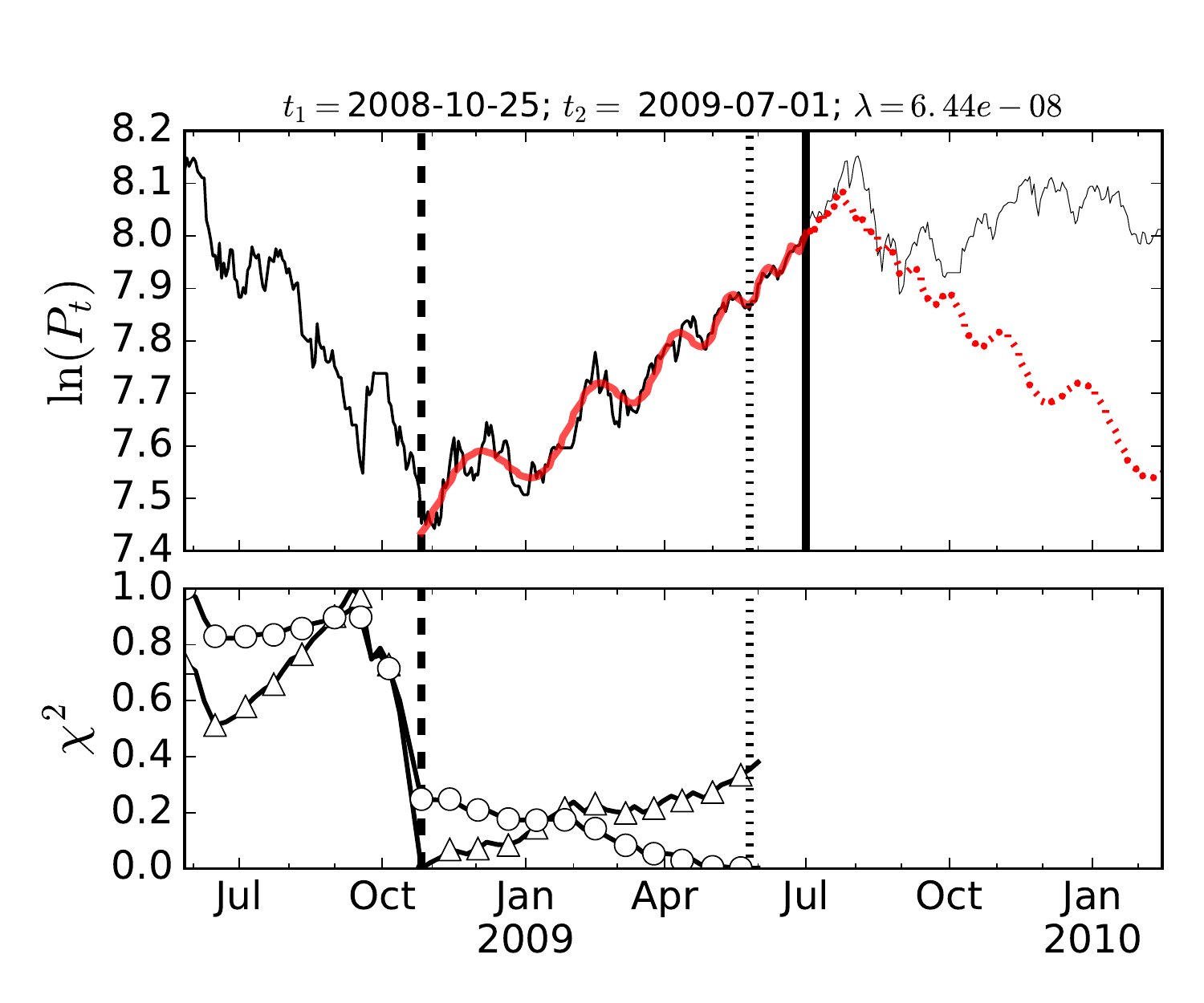}
\includegraphics[width=.4\textwidth]{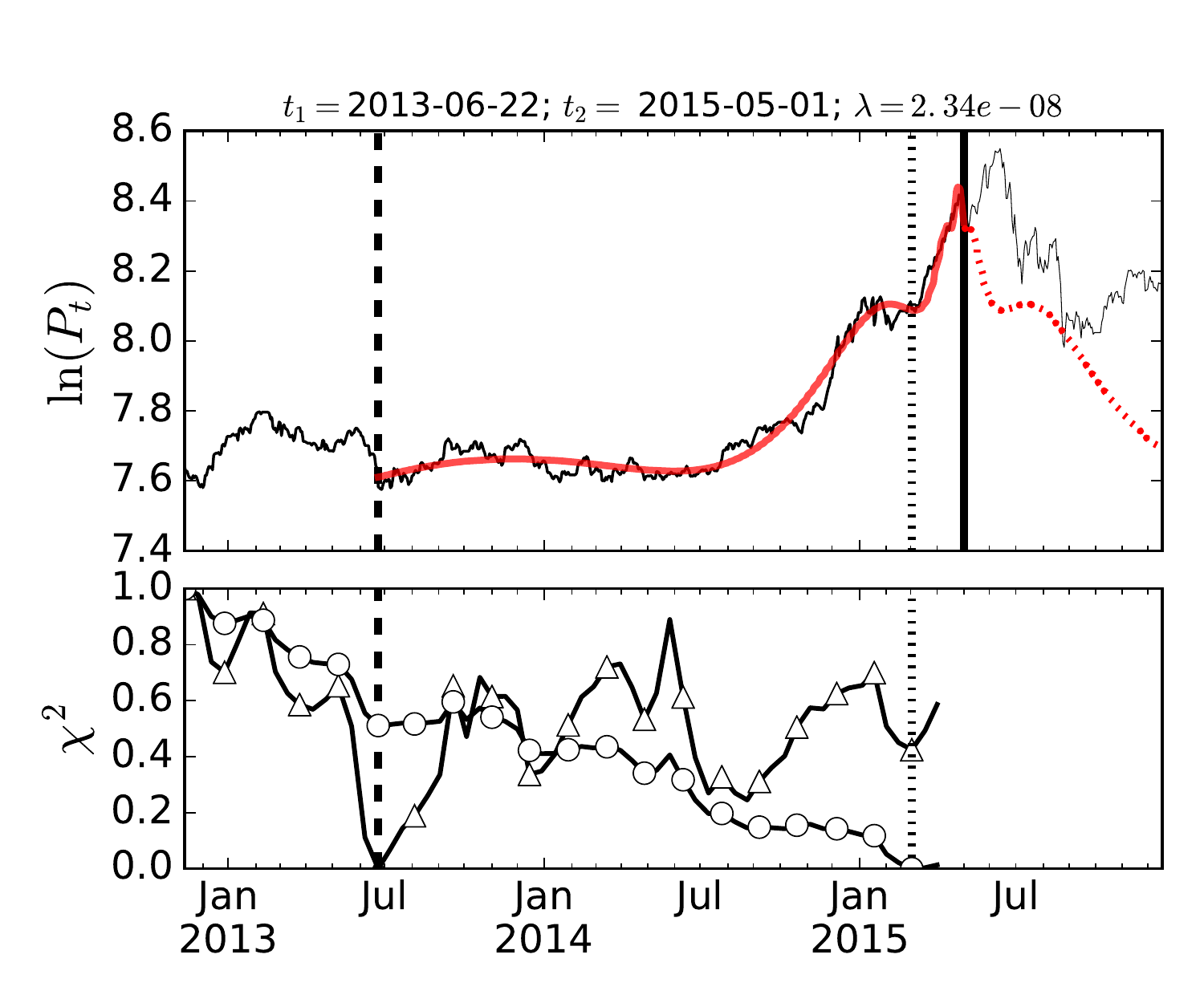}
\caption{{\bf Same as figure \ref{fig3} for the Chinese $SSEC$ index.} }\label{fig6}
\end{center}
\end{figure}

For the IBovespa and the SSEC Index (Figures (\ref{fig5}) and (\ref{fig6}) respectively), the huge superiority of the
Lagrange regularisation of the $\chi^{2}_{np}(\bm{\Phi})$ vs. the $\chi^{2}_{np}(\bm{\Phi})$ metric is again obvious. 
For each of the four chosen $t_2$'s in each figure, 
$\chi^{2}_{\lambda}(\bm{\Phi})$ exhibits a well-marked minimum corresponding to a well-defined starting time for the 
corresponding bubble. These objectively identified $t_1$ correspond pleasantly to what the eye would have chosen.
They pass the ``smell test'' \citep{Solow2010}.  In contrast, the $\chi^{2}_{np}(\bm{\Phi})$ metric provides essentially no guidance
on the determination of $t_1$.

\section{Conclusion}\label{last}

We have presented a novel goodness-of-fit metric, aimed at comparing goodnesses-of-fit across
a nested hierarchy of data sets of shrinking sizes. This is motivated by the question of 
identifying the start time of financial bubbles, but applies more generally to any calibration 
of time series in which the start time of the latest regime of interest is unknown.
We have introduced a simple and physically motivated way to correct for the overfitting bias
associated with shrinking data sets, which we refer to at the Lagrange regularisation of the 
$\chi^{2}_{np}(\bm{\Phi}) := \frac{1}{N-p} SSR$. We have suggested that the bias can be captured by a Lagrange 
regularisation parameter $\lambda$. In addition to helping remove or alleviate the bias, this parameter
can be used as a diagnostic parameter, or ``overfit measure'',
quantifying the tendency of the model to overfit the data.  
It is a function of both the specific realisation of the data and of how the model matches the
generating process of the data. 

Applying the Lagrange regularisation of the $\chi^{2}_{np}(\bm{\Phi})$ to simple linear regressions with a change point,
synthetic models of financial bubbles with a well-defined transition regime and to a number
of financial time series (US S\&P500, Brazil IBovespa and China SSEC Indices), 
we document its impressive superiority compared with the $\chi^{2}_{np}(\bm{\Phi})$ metric.
In absolute sense, the Lagrange regularisation of the $\chi^{2}_{np}(\bm{\Phi})$ is found
to provide very reasonable and well-defined determinations of the starting times 
for major bubbles such as the bubbles ending with the 1987 Black-Monday, 
the 2008 Sub-prime crisis and minor speculative bubbles on other Indexes, without any further exogenous information.

\section*{Appendix}

\clearpage
\begin{lstlisting}
// Python script for computing the Lambda regulariser metric - OLS case. 
// Copyright: G.Demos @ ETH-Zurich - Jan.2017

########################
def simulateOLS():
    """ Generate synthetic OLS as presented in the paper """ 
    nobs = 200
    X    = np.arange(0,nobs,1)
    e    = np.random.normal(0, 10, nobs)
    beta = 0.5
    Y    = [beta*X[i] + e[i] for i in range(len(X))]
    Y = np.array(Y)
    X = np.array(X)
    Y[:100] = Y[:100] + 4*e[:100]
    Y[100:200] = Y[100:200]*8
    
    return X, Y


########################
def fitDataViaOlsGetBetaAndLine(X,Y):
    """ Fit synthetic OLS """
    beta_hat = np.dot(X.T,X)**-1. * np.dot(X.T,Y) # get beta
    Y = [beta_hat*X[i] for i in range(len(X))] # generate fit
    
    return Y


########################    
def getSSE(Y, Yhat, p=1, normed=False):
    """ 
    Obtain SSE (chi^2)
    p -> No. of parameters
    Y -> Data
    Yhat -> Model
    """
    error = (Y-Yhat)**2.
    obj = np.sum(error)
    if normed == False:
        obj = np.sum(error)
    else:
        obj = 1/np.float(len(Y) - p) * np.sum(error)
        
    return obj


########################
def getSSE_and_SSEN_as_a_func_of_dt(normed=False, plot=False):
    """ Obtain SSE and SSE/N for a given shrinking fitting window  w """
    # Simulate Initial Data
    X, Y = simulateOLS()
    
    # Get a piece of it: Shrinking Window
    _sse = []
    _ssen = []
    for i in range(len(X)-10): # loop t1 until: t1 = (t2 - 10):
        xBatch = X[i:-1]
        yBatch = Y[i:-1]
        YhatBatch = fitDataViaOlsGetBetaAndLine(xBatch, yBatch)
        sse = getSSE(yBatch, YhatBatch, normed=False)
        sseN = getSSE(yBatch, YhatBatch, normed=True)
        _sse.append(sse)
        _ssen.append(sseN)

    if plot == False:
        pass
    else:
        f, ax = plt.subplots(1,1,figsize=(6,3))
        ax.plot(_sse, color='k')
        a = ax.twinx()
        a.plot(_ssen, color='b')
        plt.tight_layout()
    if normed==False:
        return _sse, _ssen, X, Y # returns results + data
    else:
        return _sse/max(_sse), _ssen/max(_ssen), X, Y # returns results + data
    
    
########################
def LagrangeMethod(sse):
    """ Obtain the Lagrange regulariser for a given SSE/N"""
    # Fit the decreasing trend of the cost function
    slope = calculate_slope_of_normed_cost(sse) 
    
    return slope[0]
    
    
########################
def calculate_slope_of_normed_cost(sse):
    #Create linear regression object using statsmodels package
    regr = linear_model.LinearRegression(fit_intercept=False)
    
    # create x range for the sse_ds
    x_sse = np.arange(len(sse))
    x_sse = x_sse.reshape(len(sse),1)
    
    # Train the model using the training sets
    res = regr.fit(x_sse, sse)
    
    return res.coef_


########################
def obtainLagrangeRegularizedNormedCost(X, Y, slope):
     """ Obtain the Lagrange regulariser for a given SSE/N Pt. III"""
    Yhat = fitDataViaOlsGetBetaAndLine(X,Y) # Get Model fit
    ssrn_reg  = getSSE(Y, Yhat, normed=True) # Classical SSE
    ssrn_lgrn = ssrn_reg - slope*len(Y) # SSE lagrange
    
    return ssrn_lgrn


########################
def GetSSEREGvectorForLagrangeMethod(X, Y, slope):
    
    """
    X and Y used for calculating the original SSEN
    slope is the beta of fitting OLS to the SSEN 
    """
    
    # Estimate the cost function pondered by lambda using a Shrinking Window.
    _ssenReg = []
    for i in range(len(X)-10):
        xBatch = X[i:-1]
        yBatch = Y[i:-1]
        regLag = obtainLagrangeRegularizedNormedCost(xBatch,
        			yBatch,
					slope)
        _ssenReg.append(regLag)
    
    return _ssenReg
   
    
\end{lstlisting}

\clearpage
\section{bibliography}
\bibliographystyle{apalike}
\bibliography{ref}

\end{document}